\documentclass[seceq]{ptptex}
\usepackage{epsfig}
\usepackage{psfrag}
\usepackage{wrapft}

\title{
Orbital order and fluctuations in Mott insulators
}

\author{
Giniyat {\sc Khaliullin} \footnote{E-mail: G.Khaliullin@fkf.mpg.de}
}

\inst{
Max-Planck-Institut~f\"{u}r~Festk\"{o}rperforschung, 
Heisenbergstr.~1, D-70569 Stuttgart, Germany}

\abst{
Basic mechanisms controlling orbital order and orbital fluctuations 
in transition metal oxides are discussed. The lattice driven classical 
orbital picture, {\it e.g.} like in manganites LaMnO$_3$, is contrasted 
to the quantum behavior of orbitals in frustrated superexchange models as  
realised in pseudocubic titanites ATiO$_3$ and vanadates AVO$_3$. 
In YVO$_3$, the lattice and superexchange effects strongly 
compete -- this explains the extreme sensitivity of magnetic states 
to temperature and doping. Lifting the $t_{2g}$ orbital degeneracy by a 
relativistic spin-orbital coupling is considered on example of the 
layered cobaltates. We find that the spin-orbital mixing of low-energy 
states leads to unusual magnetic correlations in a triangular 
lattice of the CoO$_2$ parent compound. Finally, the magnetism of sodium-rich 
compounds Na$_{1-x}$CoO$_2$ is discussed in terms of a spin/orbital 
polaronic liquid.     
}

\begin{document}

\maketitle

\section{Introduction}
It is now well recognized that the orbital degrees of freedom are an
important control parameter for physical properties of transition
metal oxides \cite{Ima98,Tok00,Mae04}. The orbital quantum number specifies 
the electron density distribution in crystal, hence providing a link 
between magnetism and structure of the chemical bonds~\cite{Goo63,Kan59}. 
When the symmetry of the crystal field experienced by a magnetoactive  
$d$-electron is high, we encounter the orbital degeneracy problem.    
In case of a single impurity this degeneracy, being an exact symmetry 
property, cannot be lifted by orbital-lattice coupling; instead, 
a purely electronic degeneracy is replaced by the same degeneracy  
of so-called vibronic states, in which the orbital dynamics and lattice 
vibrations are entangled and cannot be longer separated. 
However, the orbital degeneracy must be lifted in a dense system 
of magnetic ions. The basic physical mechanisms that quench the orbital 
degrees of freedom in a solid ---  via the orbital-lattice Jahn-Teller 
(JT) coupling, via the superexchange (SE) interactions between the orbitals, 
and via the relativistic spin-orbit coupling --- are well known and have 
extensively been discussed earlier \cite{Goo63,Kan59}; see, in particular, 
a review article \cite{Kug82}. The purpose of this paper is 
to discuss some recent developments in the field.  

Usually, it is not easy to recognize which mechanism, if any, is dominant 
in a particular material of interest. 
As a rule, one expects strong JT interactions for the 
orbitals of $e_g$-symmetry as they are directed towards the ligands. 
On the other hand, Ti, V, Ru, {\it etc.} ions with $t_{2g}$ orbital degeneracy 
are regarded as "weak Jahn-Teller" ions, and the other mechanisms listed above 
might be equally or even more important in compounds based on these ions. 
An empirical indication for the strong JT case is a spin-orbital separation 
in a sense of very different (magnetic $T_m$ and structural $T_{str}$) 
transition temperatures and large lattice distortions observed. 
Orbitals are (self)trapped by these coherent lattice distortions, and 
$d$-shell quadrupole moments are ordered regardless to whether spins are 
ordered or not. While the other two mechanisms are different in this 
respect and better characterized, loosely speaking, by spin-orbital 
confinement indicated by $T_m\simeq T_{str}$. Indeed, strong coupling 
between spin and orbital orderings/fluctuations is an intrinsic feature of 
the superexchange and spin-orbit interactions "by construction".      
The superexchange mechanism becomes increasingly effective 
near the Mott metal-insulator transition, because the intensity 
of virtual charge fluctuations 
(which are ultimately responsible for the exchange interactions)  
is large in small charge-gap materials. An increased virtual kinetic 
energy of electrons near the Mott transition --- in other words,  
the proximity to the metallic state --- can make the electronic exchange 
more efficient in lifting the orbital degeneracy than the 
electron-lattice coupling \cite{Kha00,Kha01a}. 
      
As the nature of interactions behind the above mechanisms are different, 
they usually prefer different orbital states and compete. A quantitative 
description of the orbital states is thus difficult in general but possible 
in some limiting cases. In particular, if the ground state orbital 
polarization is mainly due to strong lattice distortions, one can 
just "guess" it from the symmetry point of view, as in standart crystal 
field theory. Low-energy orbital fluctuations are suppressed in this case,  
and it is sufficient to work with spin-only Hamiltonians operating within 
the lowest classical orbital state. As an example of such a simple case, 
we consider in {\it Section 2} the magnetic and optical properties 
of LaMnO$_3$. 

The limit of strong (superexchange and relativistic spin-orbital) coupling 
is more involved theoretically, as one has to start in this case 
with a Hamiltonian which operates in the full spin-orbital Hilbert space,  
and derive the ground state spin-orbital wavefunction by 
optimizing the intersite correlations. 
It turns out that this job cannot be completed on a simple classical level; 
one realizes soon that the spin-orbital Hamiltonians possess a large 
number of classical configurations with the same energy. 
Therefore, theory must include quantum fluctuations in order to lift the 
orbital degeneracy and to select the ground state, in which the orbitals 
and spins can hardly be separated --- a different situaion to compare  
with the above case. 
The origin of such complications lies in the spatial anisotropy 
of orbital wavefunctions. This leads to a specific, non-Heisenberg form of  
the orbital interactions which are frustrated on high-symmetry 
lattices containing several equivalent bond directions.    
Physically, orbital exchange interactions on different bonds require 
the population of different orbital states and hence compete. 
This results in an infinite degeneracy of the classical ground states.  
A substantial part of this paper is devoted to illustrate how 
the orbital frustration is resolved by quantum effects 
in spin-orbital models, and to discuss results in the context 
of titanites and vanadates with perovskite structure ({\it Section 3}). 
We will also demonstrate a crucial difference in the low-energy 
behavior of superexchange models for $e_g$ and $t_{2g}$ orbitals. 

In some cases, the competition between the superexchange and orbital-lattice 
interactions result in a rich phase diagram, including mutually exclusive 
states separated by first order transitions. A nice example of this 
is YVO$_3$ discussed in {\it Section 4}. In particular, we show how 
a competition between the three most relevant spin/orbital electronic 
phases is sensitively controlled in YVO$_3$ by temperature or 
small amounts of hole-doping. 

The last part of the paper, {\it Section 5}, discusses the role of
the relativistic spin-orbit coupling $\lambda_{so}$, which effectively reduces 
the orbital degeneracy already on the single-ion level. This coupling 
mixes-up the spin and orbital quantum numbers in the ground state 
wavefunction, such that the magnetic ordering implies both the orbital 
and spin ordering at once. The spin-orbit coupling mechanism might 
be essential 
in insulating compounds of late-3$d$ ions with large $\lambda_{so}$ 
and also in 4$d$-ruthenates. We focus here on two-dimensional 
cobaltates with a triangular lattice, and present a theory which 
shows that very unusual magnetic states can be stabilized 
in CoO$_2$ planes by spin-orbit coupling. We also discuss 
in that section, how the orbital degeneracy and well known 
spin-state flexibility of cobalt ions lead to a polaron-liquid 
picture for the sodium-rich compounds Na$_{1-x}$CoO$_2$, and explain 
their magnetic properties. 

Our intention is to put more emphasis on a comparative discussion 
of different mechanisms of lifting the orbital degeneracy, by considering 
their predictions in the context of particular compounds. Apart from 
discussing previous work, the manuscript presents many original 
results which either develop known ideas or are completely new. 
These new results and predictions made may help 
to resolve controversial issues and to discriminate between the 
different models and viewpoints. 

\section{Lifting the orbital degeneracy by lattice distortions}
Let us consider the Mott insulators with composition ABO$_3$, 
where A and B sites 
accommodate the rare-earth and transition metal ions, respectively. 
They crystallize in a distorted perovskite structure \cite{Ima98,Goo04}. 
Despite a very simple, nearly cubic lattice formed by magnetic ions, 
a variety of spin structures are observed in these compounds: 
An isotropic $G$-type antiferromagnetism (AF) in LaTiO$_3$, 
isotropic ferromagnetism (F) in YTiO$_3$, and also anisotropic 
magnetic states such as $C$-type AF in LaVO$_3$ and 
$A$-type AF in LaMnO$_3$~\cite{note1}. The richness of the spin-orbital 
states, realized in these compounds with a similar lattice structure, 
already indicates that different mechanisms lifting the orbital degeneracy 
might be at work, depending on the type of orbitals, the spin value, 
and on the closeness to the Mott transition. 
 
Considering the lattice distortions in perovskites,   
one should distinguish distortions of two different origins: 
(A) The first one is due to ionic-size mismatch effects,  
that generate cooperative rotations and also some distortions 
of octahedra in order to fulfill the close-packing conditions 
within a perovskite structure. These are the "extrinsic" 
deviations from cubic symmetry, in a sense that they are not triggered by 
orbitals themselves and are present even in perovskites having no orbital 
degeneracy at all, {\it e.g.} in LaAlO$_3$ or LaFeO$_3$. The orbitals 
may split and polarize under the extrinsic deformations, but they 
play essentially the role of spectators, and to speak of 
"orbital ordering" in a sense of cooperative phenomenon (such as "spin 
ordering") would therefore be misleading in this case. 
(B) Secondly, cooperative JT-distortions, 
which are generated by orbital-lattice coupling itself at  
the orbital ordering temperature (seen by "eyes" as a structural 
transition). 

Usually, these two contributions are superimposed on each other. 
Sometimes, it is not easy to identify which one is dominant. 
The temperature dependence of the distortion is helpful in this context.   
Manganites show an orbital order-disorder phase transition at high 
temperatures at about 800 K, below which a large, of the order of 15\%, 
distortion of octahedra sets in. This indicates the dominant role 
of a cooperative JT physics, which is natural for $e_g$ orbital systems 
with strong coupling between the oxygen vibrations and $e_g$ quadrupole. 
In contrast, no {\it cooperative} structural phase transition 
has thus far been observed in titanates, suggesting that small 
lattice distortions present 
in these compounds are mostly due to the ionic-size mismatch effects. 
This seems also plausible, as $t_{2g}$-orbital JT coupling is weak.   
 
Whatever the origin, the lattice distortions generate 
low (noncubic) symmetry components in the crystal field potential,  
which split the initially degenerate orbital levels. While structure 
of the lowest, occupied crystal-field level can be determined without much 
problem --- simply by symmetry considerations, the level splittings 
are very sensitive to the value of ({\it i}) distortions and ({\it ii}) the 
orbital-lattice coupling (both factors being much smaller in $t_{2g}$ 
orbital compounds). If the splittings are large enough 
to localize electrons in the lowest crystal-field level and to suppress 
the intersite orbital fluctuations, classical treatment of 
orbitals is justified. Accordingly, the $e_g$-electron density 
is determined by a function parameterized via the site-dependent 
classical variables $\alpha_i$:     
\begin{equation}
\label{1angle}
\psi_i=\cos\alpha_i|3z^2-r^2>+\sin\alpha_i|x^2-y^2>\;,  
\end{equation}
while the occupied $t_{2g}$ orbital can be expressed via the two angles:
\begin{equation}
\label{2angle}
\psi_i=
\cos\alpha_i|xy>+\sin\alpha_i\cos\beta_i|yz>+\sin\alpha_i\sin\beta_i|zx>.  
\end{equation}
A knowledge of these functions allows one to express a various experimental 
observables such as spin interactions, optical absorption and Raman scattering 
intensities, {\it etc.} in terms of a few "orbital angles". [In general, 
classical orbital states may include also a complex wave-functions, 
but they are excluded in the case of strong lattice 
distortions\cite{Goo63,Kug82}]. These angles can thus be determined 
from the experimental data, and then compared with those 
suggested by a crystal-field theory. Such a classical approach 
is a great simplification of the "orbital physics", and it has widely 
and successfully been used in the past. Concerning the orbital excitations 
in this picture, they can basically be regarded as a nearly localized 
transitions between the crystal-field levels. We demonstrate below how 
nicely this canonical "orbital-angle" scenario does work in manganites, 
and discuss its shortcomings in titanites and vanadates. 
 
\subsection{$e_g$ orbitals: The case of LaMnO$_3$} 
Below a cooperative JT phase transition at $\sim$800~K$\gg T_N$, 
a two-sublattice orbital order sets in. 
Suggested by C-type arrangement of octahedron elongation 
(staggered within $ab$ planes and repeated along $c$ direction), the lowest 
crystal field level can be parameterized via the orbital 
angle $\theta$ \cite{Kan59} as follows: 
\begin{equation}
\label{theta}
|\pm>=\cos\frac{\theta}{2}|3z^2-r^2>\pm\sin\frac{\theta}{2}|x^2-y^2>. 
\end{equation}  
In this state, the $e_g$ electron distribution is spatially asymmetric, 
which leads to strong anisotropy in spin exchange couplings and optical 
transition amplitudes. Thus, the orbital angle $\theta$ can be 
determined from a related experiments and compared with 
$\theta\sim 108^{\circ}$ \cite{Kan59} suggested by structural data. 

{\it Spin interactions}.--- For manganites, there are several intersite 
virtual transitions $d^4_id^4_j\rightarrow d^3_id^5_j$, each contributing to 
the spin-orbital superexchange. 
The corresponding energies can be parameterized via the Coulomb 
repulsion $U=A+4B+3C$ for electrons residing in the same $e_g$ orbital 
($Un_{\alpha\uparrow}n_{\alpha\downarrow}$), the Hund's integral 
$J_H=4B+C$ between $e_g$ spins in different orbitals 
($-2J_H\vec s_{\alpha} \cdot \vec s_{\beta}$), the Hund's integral 
$J'_H=2B+C$ between the $e_g$ spin and the $t_{2g}$-core spin 
($-2J'_H\vec S_t \cdot \vec s_e$), and, finally, the 
Jahn-Teller splitting $\Delta_{JT}$ between different $e_g$-orbitals: 
$\Delta_{JT}(n_{\beta}-n_{\alpha})/2$.  
($A,B,C$ are the Racah parameters, see Ref.~\cite{Gri61} for details). 
For the $e_g$ electron hopping $d^4_id^4_j\rightarrow d^3_id^5_j$, 
we obtain the following five different transition energies: 
\begin{eqnarray}
\label{En}
E_1&=&U+\Delta_{JT}-3J_H~, \\ \nonumber 
E_2&=&U+\Delta_{JT}-3J_H+5J'_H~, \\ \nonumber 
E_3&=&U+\Delta_{JT}+3J'_H-\sqrt{\Delta_{JT}^2+J_H^2}~, \\ \nonumber 
E_4&=&U+\Delta_{JT}+3J'_H-J_H~, \\ \nonumber 
E_5&=&U+\Delta_{JT}+3J'_H+\sqrt{\Delta_{JT}^2+J_H^2}~. 
\end{eqnarray}
There are also (weaker) transitions associated with $t_{2g}$ electron 
hoppings, which provide an isotropic AF coupling between the spins  
$J_t(\vec S_i\cdot\vec S_j)$, with $J_t=t'^2/2E_t$ and $S=2$ of 
Mn$^{3+}$. Here, $E_t=U+J_H+2J'_H$ is an average excitation energy for the 
intersite $t_{2g}-$electron hoppings, and $t'\simeq t/3$ follows 
from the Slater-Koster relation. 

From a fit to the optical conductivity in LaMnO$_3$, the energies $E_n$ 
have been obtained~\cite{Kov04}. Then, it follows from Eqs.(\ref{En}) that 
$U\sim 3.4$~eV, $J'_H\sim 0.5$~eV, $J_H\sim 0.7$~eV and 
$\Delta_{JT}\sim 0.7$~eV (in the present notations for $U$, $J'_H$, $J_H$ 
given above; Ref.~\cite{Kov04} used instead $\tilde U =A-2B+3C$ and 
$\tilde J_H =2B+C$). The Hund's integrals are somewhat reduced from 
the atomic values $J'_H=0.65$~eV and $J_H=0.89$~eV~\cite{Gri61}. 
The $e_g$-orbital splitting is substantial, suggesting that orbitals  
are indeed strongly polarized by static JT distortions, and justifying 
a crystal-field approach below 800~K. A large JT binding 
energy ($=\Delta_{JT}/4$) also indicates that dynamical distortions 
of octahedra are well present above 800~K, thus a structural 
transition is of the order-disorder type for these distortions 
and $e_g-$quadrupole moments. This is in full accord with a view 
expressed in Ref.\cite{Mil96}. Note, however, that JT energy scales are 
smaller than those set by correlations, thus LaMnO$_3$ has to be regarded 
as a typical Mott insulator.   

The superexchange Hamiltonian $H$ consists of several terms, 
$H_n^{(\gamma)}$, originating from virtual hoppings 
with energies $E_n$ (\ref{En}) in the intermediate state ($\gamma$ denotes 
the bond directions $a,b,c$). 
Following the derivation of Ref.~\cite{Fei99}, we obtain:   
\begin{eqnarray}
\label{Hn}
H_{ij}^{(\gamma)}=\frac{t^2}{20}\big[
&-&\frac{1}{E_1}(\vec S_i\cdot\vec S_j+6)(1-4\tau_i\tau_j)^{(\gamma)} 
+(\frac{3}{8E_2}+\frac{5}{8E_4})(\vec S_i\cdot\vec S_j-4)
(1-4\tau_i\tau_j)^{(\gamma)} \nonumber \\
&+&(\frac{5}{8E_3}+\frac{5}{8E_5})(\vec S_i\cdot\vec S_j-4)
(1-2\tau_i)^{(\gamma)}(1-2\tau_j)^{(\gamma)}\big],  
\end{eqnarray}
where the $e_g$-pseudospins $\tau^{(\gamma)}$ are defined as 
$2\tau_i^{(c)}=\sigma_i^{z}$, 
$2\tau_i^{(a)}=\cos\phi\sigma_i^{z}+\sin\phi\sigma_i^{x}$,  
$2\tau_i^{(b)}=\cos\phi\sigma_i^{z}-\sin\phi\sigma_i^{x}$ 
with $\phi=2\pi/3$. Here, $\sigma^z$ and $\sigma^x$ are the Pauli matrices, 
and $\sigma^z$=1 (-1) corresponds to a $|x^2-y^2\rangle$ ($|3z^2-r^2\rangle$) 
state. Note that no $\sigma^y$-component is involved in $\tau^{(\gamma)}$ 
operators; this relfects the fact that the orbital angular momentum is fully 
quenched within the $e_g$ doublet. Physically, a pseudospin 
$\tau^{(\gamma)}-$structure of (\ref{Hn}) reflects 
the dependence of spin interactions 
on the orbitals that are occupied, expressing thereby a 
famous Goodenough-Kanamori rules in a compressed formal way \cite{Kug82}. 

In a classical orbital state (\ref{theta}), operators 
$\sigma^{\alpha}$ can be regarded as a numbers $\sigma^z=-\cos\theta$ 
and $\sigma^x=\pm\sin\theta$, thus we may set 
$(1-4\tau_i\tau_j)^{(c)}=\sin^2\theta$, 
$(1-4\tau_i\tau_j)^{(ab)}=(3/4+\sin^2\theta)$ {\it etc.} in Eq.(\ref{Hn}). 
At this point, the spin-orbital model (\ref{Hn}) "collapses" into 
the Heisenberg 
Hamiltonian, $J_{\gamma}(\theta)(\vec S_i\cdot\vec S_j)$, and only "memory" 
of orbitals that is left is in a possible anisotropy of spin-exchange 
constants $J_{a,b,c}$:  
\begin{eqnarray}
\label{J}
J_c&=&\frac{t^2}{20}\big[\big(-\frac{1}{E_1}+\frac{3}{8E_2}+
\frac{5}{8E_4}\big)\sin^2{\theta}+\frac{5}{8}
\big(\frac{1}{E_3}+\frac{1}{E_5}\big)(1+\cos{\theta})^2\big]+J_t~, \\
J_{ab}&=&\frac{t^2}{20}\big[\big(-\frac{1}{E_1}+\frac{3}{8E_2}+
\frac{5}{8E_4}\big)\big(\frac{3}{4}+\sin^2{\theta}\big)+
\frac{5}{8}\big(\frac{1}{E_3}+\frac{1}{E_5}\big)
\big(\frac{1}{2}-\cos{\theta}\big)^2\big]+J_t.
\nonumber
\end{eqnarray}
Note that the $E_1$--term has a negative (ferromagnetic) sign. 
This corresponds to the high-spin intersite transition with the 
lowest energy $E_1$, which shows up in optical absorption spectra 
as a separate band near 2 eV \cite{Kov04}. From the spectral 
weight of this line, one can determine the value of $t$, see below. 
All the other terms come from low-spin transitions. The orbital 
angle $\theta$ controls the competition between a ferromagnetic $E_1-$term 
and AF $E_2$,...,$E_5-$contributions in a bond-selective way.  

{\it Optical intensities}.---Charge transitions 
$d^4_id^4_j\rightarrow d^3_id^5_j$ are optically active. 
The spectral shape of corresponding lines are controlled by band motion 
of the excited doublons (holes) in the upper (lower) Hubbard bands and 
by their excitonic binding effects. The intensity of each line at $E_n$ is 
determined by a virtual kinetic energy, and thus, according to the optical 
sum rule, can be expressed via the expectation values of superexchange terms  
$H_n^{(\gamma)}(ij)$ for each bond direction $\gamma$. 

The optical intensity data is often quantified via the effective 
carrier number, 
\begin{equation}
N_{eff,n}^{(\gamma)}= \frac{2m_0v_0}{\pi e^2}
\int_0^{\infty}\sigma_n^{(\gamma)}(\omega)d\omega, 
\end{equation}
where $m_0$ is the free electron mass, and $v_0=a_0^3$ 
is the volume per magnetic ion. Via the optical sum rule applied to a given
transition with $n=1,...,5$, the value of $N_{eff,n}^{(\gamma)}$ 
can be expressed as follows \cite{Kha04a}:  
\begin{equation}
\label{sumrule}
N_{eff,n}^{(\gamma)}=\frac{m_0a_0^2}{\hbar^2}K_n^{(\gamma)}
=-\frac{m_0a_0^2}{\hbar^2}\big\langle 2H_n^{(\gamma)}(ij)\big\rangle.
\end{equation}
Here, $K_n^{(\gamma)}$ is the kinetic energy gain associated with a given 
virtual transition $n$ for a bond $\langle ij\rangle$ along axis $\gamma$. 
The second equality in this expression states that $K_n^{(\gamma)}$, 
hence the intensity of a given optical transition, is controlled by 
expectation value of the corresponding term $H_n^{(\gamma)}$ 
in spin-orbital superexchange interaction (\ref{Hn}). Via the operators 
$\tau^{(\gamma)}$, each optical transition $E_n$ obtains its own 
dependence on the orbital angle $\theta$.  
Thus, Eq.(\ref{sumrule}) forms a basis for the study of orbital states 
by means of optical spectroscopy, in addition to the magnetic data. 
Using a full set of the optical and magnetic data, it becomes possible 
to quantify more reliably the values of $t$, $U$, $J_H$ in crystal, and to 
determine the SE-energy scale, as it has been proposed in Ref.~\cite{Kha04a}, 
and worked out specifically for LaVO$_3$~\cite{Kha04a} and 
LaMnO$_3$~\cite{Kov04}. 

Physically, spin and orbital correlations determine the optical 
intensities of different transitions $E_n$ via the selection rules, 
which are implicit in Eq.(\ref{Hn}). For instance, the intensity of the 
high-spin transitions obtained from the $E_1$-term in Eq.(\ref{Hn}) reads:  
\begin{eqnarray}
\label{Kc}
K^{(c)}_1&=&\frac{t^2}{10E_1}<\vec S_i\cdot\vec S_j+6>^{(c)}\sin^2\theta, 
\\ \nonumber
K^{(ab)}_1&=&\frac{t^2}{10E_1}<\vec S_i\cdot\vec S_j+6>^{(ab)}
(3/4+\sin^2\theta).
\end{eqnarray}
At $T \ll T_N\sim 140$K, $<\vec S_i\cdot\vec S_j >^{(ab)} \Rightarrow$~4 and 
$<\vec S_i\cdot\vec S_j>^{(c)}\Rightarrow$~$-$4 for the $A$-type classical 
spin state, while $<\vec S_i\cdot\vec S_j>\Rightarrow 0$ at $T \gg T_N$. 
Thus, both spin and orbital correlations induce a strong polarization
and temperature dependence of the optical conductivity. Note, that the 
$e_g$-hopping integral $t$ determines the overall intensity; we find that 
$t\simeq 0.4$~eV fits well the observed optical weights~\cite{Kov04}.  
This gives $4t^2/U\simeq 190$~meV (larger than in cuprates, since $t$ stands 
here for the $3z^2-r^2$ orbitals and thus is larger than 
a planar $x^2-y^2$ orbital-transfer). 

A quantitative predictions of the above "orbital-angle" theory for 
spin-exchange constants and optical intensities, expressed 
in Eqs.(\ref{J}) and (\ref{Kc}) solely in terms of a classical 
variable $\theta$, have nicely been confirmed in Ref.\cite{Kov04}. 
It is found that the angle $\theta=102^{\circ}$ well describes 
the optical anisotropy, and gives the exchange integrals 
$J_c=1.2$~meV, $J_{ab}=-1.6$~meV in perfect agreement with magnon 
data \cite{Hir96}. Given that "the lattice angle" $108^{\circ}$ has 
been estimated from the octahedron distortions alone, and thus may 
slightly change in reality by GdFeO$_3$-type distortions and exchange 
interactions, one may speak of a quantitative description of the entire 
set of data in a self-contained manner (everything is taken from the 
experiment). This implies the dominant role of (JT) lattice distortions 
in lifting the orbital degeneracy in manganites as expected. Of course, 
the situation changes if one injects a mobile holes, or drives a system closer 
to the Mott transition. The orbital order is indeed suppressed in LaMnO$_3$ 
under high pressure \cite{Loa01}, in spite the insulating state is 
still retained. Pressure reduces the ionic-size mismatch 
effects, and, more importantly, decreases the charge gap and thus 
enhances the kinetic energy. The latter implies an increased role of 
superexchange interactions which, as discussed later, are strongly 
frustrated on cubic lattices; consequently, a classical orbital
order is degraded. In addition, a weak $e_g-$density modulation like in case
of CaFeO$_3$\cite{Ima98} may also contribute to the orbital quenching near the 
metal-insulator transition. 

\subsection{$t_{2g}$ orbitals: Titanates}
Depending on the A-site ion radius, two types of magnetic orderings have 
thus far been identified in ATiO$_3$ compounds: (I) $G$-type AF state 
as in LaTiO$_3$, which changes to (II) ferromagnetic ordering for the smaller 
size A-ions as in YTiO$_3$, with strong suppression of transition temperatures 
in between \cite{Kat97}. Surprisingly enough, both AF and F states 
are isotropic in a sense of equal exchange couplings 
in all three crystallographic directions \cite{Kei00,Ulr02}. 
Such a robust isotropy of spin interactions (despite the fact that 
the $c$-direction bonds are structurally not equivalent to those in 
the $ab$ plane), and a kind of "soft" (instead of strong first order) 
transition between such a incompatible spin states is quite unusual. 

Another unique feature of titanites is that, different from many oxides, 
no cooperative structural phase transition is observed so far  
in LaTiO$_3$ and YTiO$_3$. (Except for a sizable magnetostriction  
in LaTiO$_3$ \cite{Cwi03}, indicating the presence of low-energy 
orbital degeneracy). This is very unusual because the JT physics could  
apparently be anticipated for Ti$^{3+}$ ions in nearly octahedral 
environment. One way of thinking of this paradox is that 
\cite{Kei00,Kha00,Kha01a,Ulr02,Kha02,Kha03}  
the titanites are located close to the Mott transition, and the enhanced 
virtual charge fluctuations, represented usually in the form of 
spin-orbital superexchange interactions, frustrate and suppress 
expected JT-orbital order (see next section). 
Yet another explanation is possible \cite{Moc03,Cwi03}, 
which is based on local crystal field picture. In this view, the orbital 
state is controlled by lattice, instead of superexchange interactions. 
Namely, it is thought that extrinsic lattice distortions caused by 
ionic-size mismatch effects (so-called GdFeO$_3$-type distortions) remove 
the orbital degeneracy for all temperatures, thus orbital dynamics 
is suppressed and no Jahn-Teller instability can develop, either. 
One caveat in this reasoning, though, is that a cooperative orbital  
transitions are {\it not prevented} by GdFeO$_3$-distortions (of the 
similar size) in other $t_{2g}$ compounds, {\it e.g.} in pseudocubic 
vanadates. Thus, the titanites seem to be unique and require other than the 
"GdFeO$_3$-explanation" for the orbital quenching. 
Nonetheless, let us consider now in more detail the predictions of such  
a crystal-field scenario, that is, the "orbital-angle" 
picture expressed in Eq.(\ref{2angle}).  
 
The perovskite structure is rather "tolerant" and can accommodate 
a variation of the A-site ionic radius by cooperative rotations 
of octahedra. This is accompanied by a shift of A-cations such that 
the distances A--O and A--Ti vary somewhat. 
Also the oxygen octahedra get distorted, but their distortion ($\sim$ 3\%) 
is far less than that of the oxygen coordination around the A-cation.       
The nonequivalence of the Ti--A distances and weak distortions of
octahedra induce then a noncubic crystal field for the Ti-ion. 
In LaTiO$_3$, the octahedron is nearly perfect, but the Ti--La distance 
along one of the body diagonals is shorter\cite{Cwi03}, suggesting the 
lowest crystal-field level of $\sim |xz+yz\pm xy\rangle/\sqrt3$ 
symmetry, as has been confirmed in Refs.\cite{Moc03,Cwi03} 
by explicit crystal-field calculations. This state describes  
an orbital elongated along the short La--Ti--La diagonal, and the 
sign $(\pm)$ accounts for the fact that the short diagonal 
direction alternates along the $c$-axis (reflecting 
a mirror plane present in crystal structure). 
Thus, lattice distortions impose the orbital pattern of $A$-type structure. 
In YTiO$_3$, the A-site shifts and concomitant distortions of octahedra, 
{\it induced} \cite{Pav04} by ionic-size mismatch effects including A--O and 
A--Ti covalency, are stronger. 
A crystal-field \cite{Moc03,Cwi03} and band structure \cite{Miz96,Saw97} 
calculations predict that these distortions stabilize a four-sublattice 
pattern of planar orbitals 
\begin{equation}
\label{function}
\psi_{1,3}\sim|xz\pm xy\rangle/\sqrt2 \;\;\;\; {\rm and} \;\;\;\;
\psi_{2,4}\sim|yz\pm xy\rangle/\sqrt2~.
\end{equation} 
Note that $xy$ orbitals 
are more populated (with two times larger filling-factor when averaged 
over sites). This state would therefore lead to a sizable anisotropy of
the electronic properties. We recall that all the above orbital patterns are 
nearly temperature independent, since {\it no cooperative} JT structural 
transition (like in manganites) is involved here. 

Earlier neutron diffraction \cite{Aki01} and NMR \cite{Kiy03} experiments 
gave support to the crystal-field ground states described above; however, 
this has recently been reconsidered in Ref.\cite{Kiy05}. Namely, 
NMR-spectra in YTiO$_3$ show that the $t_{2g}$-quadrupole moment 
is in fact markedly reduced from the predictions of crystal-field theory, 
and this has been attributed to quantum orbital fluctuations \cite{Kiy05}. 
On the other hand, this maybe not easy to reconcile with the earlier 
"no orbital-fluctuations" interpretation of NMR spectra in 
{\it less distorted} LaTiO$_3$~\cite{Kiy03}, --- thus no final 
conclusion can be made at present from NMR data alone~\cite{Kub05}. 

For the {\it first-principle} study of effects of lattice distortions 
on the orbital states in titanites, we refer to the recent 
work~\cite{Pav05} in which: the ground-state orbital polarization, 
(crystal-field) energy level structure, and spin exchange interactions 
have been calculated within the LDA+DMFT framework. In this study, 
which attempts to combine a quantum chemistry with strong correlations, 
a nearly complete condensation of electrons in a particular 
single orbital (an elongated, 
of the $a_{1g}$ type in LaTiO$_3$ and a planar one in YTiO$_3$, as just 
discussed) is found, in accord with a simpler crystal-field calculations 
of Refs.\cite{Moc03,Cwi03}. 
Other (empty) orbital levels are found to be located at energies 
as high as $\sim 200$~meV; this qualitatively agrees again with 
Refs.\cite{Cwi03,Moc03}, but deviates from the predictions 
of yet another first-principle study of Ref.\cite{Sol04}.     
The calculations of Ref.\cite{Pav05} give correct spin ordering patterns 
for both compounds (for a first time, to our knowledge; --- usually, it is 
difficult to get them both right simultaneously, see Ref.\cite{Sol04}). 
However, the calculated spin exchange constants:  
$J_c=-0.5$~meV, $J_{ab}=-4.0$~meV for YTiO$_3$, and 
$J_c=5.0$~meV, $J_{ab}=3.2$~meV for LaTiO$_3$, --- as well as their 
anisotropy ratio $J_c/J_{ab}$ are quite different from 
the observed values: $J_c\simeq J_{ab}\simeq -3$~meV (YTiO$_3$ \cite{Ulr02}) 
and $J_c\simeq J_{ab}\simeq 15.5$~meV (LaTiO$_3$ \cite{Kei00}). Large 
anisotropy of the calculated spin couplings $J_c/J_{ab}\sim 0.1$ in YTiO$_3$ 
is particularly disturbing; the same trend is found in Ref.\cite{Sol04}: 
$J_c/J_{ab}\sim 0.3$. The origin of anisotropy is --- put it 
simple way --- that the lattice distortions 
make the $c$ axis structurally different from the two others, and this
translates into $J_c\neq J_{ab}$ within a crystal-field picture. 
In principle, one can achieve $J_c/J_{ab}=1$ by tuning the orbital angles 
and $J_H$ by "hand", but, as demonstrated in Ref.\cite{Ulr02}, this is highly 
sensitive procedure and seems more like as an "accident" rather than 
explanation. Given that the degree of lattice distortions 
in LaTiO$_3$ and YTiO$_3$ are different\cite{Cwi03} --- as reflected 
clearly in their very different crystal-field states --- the observation 
$J_c/J_{ab}\simeq 1$ in {\it both of them} is enigmatic, and 
the results of first-principle calculations make the case complete. 

A way to overcome this problem is --- as proposed in 
Refs.\cite{Kei00,Ulr02} on physical grounds --- to relax the orbital 
polarization: this would make the spin distribution more isotropic. 
In other words, we abandon the fixed orbital states like (\ref{function}), 
and let the orbital angles to fluctuate, as they do in case of a dynamical 
JT problem. However, a dynamical aspects of JT physics as well as the
intersite superexchange fluctuations are suppressed by a static 
crystal-field splittings, which, according to Ref.\cite{Pav04}, are 
large: the $t_{2g}$ level structure reads as [0, 200, 330]~meV 
in YTiO$_3$, and [0, 140, 200]~meV in LaTiO$_3$. 

Such a strong noncubic splitting of $t_{2g}$ triplet, exceeding 
10\% of the cubic $10Dq\sim 1.5-2.0$~eV is quite surprising, 
and, in our view, is overestimated in the calculations. Indeed, 
$e_g$ orbital splitting in manganites has been found $\sim 0.7$~eV 
(see above) which is about 30\% of $10Dq$; this sounds reasonable 
because of ({\it i}) very large octahedron distortions ($\sim 15$\%) 
that ({\it ii}) strongly couple to the $e_g$-quadrupole. Both these 
factors are much smaller in titanates: ({\it i}) octahedron distortions 
are about 3\% only, while effects of further neighbors should be screened 
quite well due to a proximity of metallicity; ({\it ii}) coupling to the 
$t_{2g}$-quadrupole is much less than that for $e_g$ --- otherwise 
a cooperative JT transition would take place followed by a strong 
distortions like in manganites. Putting these together, 
we "estimate" that a $t_{2g}$ splitting should be at least by 
an order of value less than that seen for $e_g$ level 
in manganites, --- that is, well below 100 meV. This would make it possible
that orbitals start talking to spins and fluctuate as suggested in 
Ref.\cite{Kha00}. 

In general, a quantitative calculation of the level splittings in a solid 
is a rather sensitive matter (nontrivial even for well localized 
$f$-electrons). Concerning a numerical calculations, we tend to 
speculate that a notion of "orbital splitting" (as used in the models) 
might be not that well defined in first-principle calculations, 
which are designed to obtain the ground state properties 
(but not excitation spectra). This might be a particularly delicate matter 
in a strongly correlated systems, where effective orbital splittings may 
have strong $t/U$ and frequency dependences \cite{Kel04}. 
As far as a simple crystal-field
calculations are concerned, they are useful for symmetry analyses but, 
quantitatively, often strongly deviate from what is observed. 
As an example of this sort, we refer to the case of Ti$^{3+}$ ions 
in a Al$_2$O$_3$ host, where two optical transitions within a $t_{2g}$ 
multiplet have directly been observed \cite{Nel67}. The level 
splitting $\sim 13$~meV found is astonishingly small, given a sizable 
(more than 5\%) distortion of octahedron. A strong reduction of the 
level splitting from a static crystal-field model has been attributed 
to the orbital dynamics (due to the dynamical JT effect, in that case). 

As a possible test of a crystal-field calculations in perovskites, 
it should be very instructive to perform a similar work for LaVO$_3$ and 
YVO$_3$ at room temperature. The point is that there is 
{\it a cooperative, of second order} 
orbital phase transition in vanadates at low-temperature (below 200~K). 
To be consistent with this observation, the level-spacings 
induced by GdFeO$_3$-type {\it etc.} distortions (of the similar size
as in titanites) must come out very small. 

Whatever the scenario, the orbital excitations obtain a characteristic 
energy scale set by the most relevant interactions. However, the
predictions of a local crystal-field theory and a cooperative SE interactions 
for the momentum and polarization dependences of orbital response 
are radically different. 
Such a test case has recently been provided by the Raman light scattering 
from orbital excitations in LaTiO$_3$ and YTiO$_3$, see Ref.\cite{Ulr05}. 
In short, it was found that the polarization rules, dictated 
by a crystal-field theory, are in complete disagreement with those 
observed. Instead, the intensity of the Raman scattering obeys the 
cubic symmetry in both compounds when the sample is rotated. Altogether, 
magnon and Raman data reveal a universal (cubic) symmetry 
of the spin and orbital responses in titanites, which is seemingly 
not sensitive to the differences in their lattice distortions. 
Such a universality is hard to explain in terms of a crystal-field 
picture based --- by its very meaning --- on the deviations 
away cubic symmetry. 
 
Moreover, a picture based on a Heisenberg spins residing on a fully polarized 
single orbital has a problem in explaining the anomalous spin reduction in 
LaTiO$_3$~\cite{Kei00}. The magnetic moment measured is as small as 
$M\simeq 0.5 \mu_B$, so the deviation from the nearest-neighbor 
3D Heisenberg model value $M_H \simeq 0.84 \mu_B$ is unusually large:   
$\delta M/M_H = (M_H-M)/M_H \sim 40 \%$. 
As a first attempt to cure this problem, one may notice that   
the Heisenberg spin system is itself derived from the Hubbard model 
by integrating out virtual charge fluctuations (empty and doubly occupied 
states). Therefore, the amplitude of the physical magnetic moment is reduced 
from that of the low-energy Heisenberg model via $M_{ph}=M_H(1-n_h-n_d)$. 
However, this reduction is in fact very small, as one can see from the 
following argument. Let us discard for a moment the orbital fluctuations, 
and consider a single orbital model. By second-order perturbation 
theory, densities of the virtual holons $n_h$ and doublons $n_d$, 
generated by electron hoppings in a Mott insulator, can be estimated 
as $n_h=n_d \simeq z(\frac{t}{U})^2$, where $z$ is the nearest-neighbor 
coordination number. Thus, the associated moment reduction is 
$\delta M/M_H\simeq \frac{1}{2z}(\frac{2zt}{U})^2$. Even near 
the metal-insulator transition, that is for $U\simeq 2zt$, this gives 
in 3D cubic lattice $\delta M/M_H\simeq\frac{1}{2z}\simeq 8\%$ only. 
(We note that this simple consideration is supported by 2D Hubbard model 
calculations of Ref.\cite{Shi95}: the deviation of the staggered 
moment from 2D Heisenberg limit was found $\delta M/M_H\simeq 12\%$ 
at $U\simeq 8t$, in perfect agreement with the above $1/2z$--scaling.)  
This implies that the anomalous moment reduction in LaTiO$_3$ 
requires some extra physics not contained in a single-orbital 
Hubbard or Heisenberg models. (In fact, an "extra" moment reduction 
is always present whenever orbital fluctuations are suspected: 
LaTiO$_3$, LaVO$_3$ and $C$-phase of YVO$_3$). Enhanced spin fluctuations, 
a spatial isotropy of the spin exchange integrals, and similar cubic 
symmetry found in the Raman scattering from orbital fluctuations 
strongly suggest multiband effects, that is, the presence of orbital 
quantum dynamics in the ground state of titanites. 

\subsection{$t_{2g}$ orbitals: Vanadates}
In vanadium oxides AVO$_3$ one is dealing with a $t^2_{2g}$ configuration, 
obtained by a removal of an electron from the $t^3_{2g}$ shell which is 
orbitally-nondegenerate (because of the Hund's rule). From a viewpoint 
of the JT physics, this represents a hole-analogy of titanites: One hole 
in the $t_{2g}$ manifold instead of one electron in ATiO$_3$. 
Similar to the titanites, they crystallize in a perovskite structure, 
with GdFeO$_3$-type distortions increasing from La- towards Y-based 
compounds, as usual. At lower temperatures, vanadates undergo 
a cooperative structural transition (of second-order, typically). 
This very fact indicates the presence of unquenched low-energy orbital 
degeneracy, suggesting that underlying GdFeO$_3$-type distortions are 
indeed not sufficient to remove it. The structural transition temperature 
$T_{str}$ is nearly confined to the magnetic one $T_N\sim 140$~K in 
LaVO$_3$ --- the relation $T_{str}\sim T_N$ holds 
even in a hole-doped samples~\cite{Miy00}, --- while $T_{str}\sim 200$~K 
in YVO$_3$ is quite separated from $T_N\sim 120$~K. 
It is accepted that the $xy$ orbital is static below $T_{str}$ and 
accommodates one of the two magnetic electrons. An empirical 
correlation between $T_{str}$ and the A-site ionic size 
suggests that GdFeO$_3$-type distortions "help" this stabilization. 

However, a mechanism lifting the degeneracy of $xz/yz$ doublet is
controversial: Is it also of lattice origin, or controlled by 
the superexchange? Based on distortions of octahedra (albeit as 
small as in YTiO$_3$), many researchers draw a pattern of staggered 
orbitals polarized by JT interactions. This way of thinking is 
conceptually the same as in 
manganites (the only difference is the energy scales): a cooperative, 
three-dimensional JT-ordering of $xz/yz$ doublets. 
However, this leaves open the questions: Why is the JT mechanism 
so effective for the $t_{2g}$-hole of vanadates, while being apparently 
innocent (no structural transition) in titanites with 
one $t_{2g}$-electron? Why is the $T_{str}$, {\it if} driven by JT physics, 
so closely correlated with $T_N$ in LaVO$_3$? Motivated by these 
basic questions, we proposed a while ago a different view \cite{Kha01b}, 
which attributes the difference between vanadates and titanites to the 
different spin values: classical $S=1$ versus more quantum $S=1/2$. Being 
not much important for JT-physics, the spin-value is of key importance 
for the superexchange mechanism of lifting the orbital degeneracy. An apparent 
correlation between $T_{str}$ and $T_N$ in LaVO$_3$, the origin of 
$C$-type spin pattern (different from titanites), {\it etc.},  
all find a coherent explanation within the model of Ref.\cite{Kha01b}. 
This theory predicts that the $xy$ orbital becomes indeed 
classical (as commonly believed) below $T_{str}$, but, different 
from the JT scenario, $xz/yz$ doublets keep fluctuating and 
their degeneracy is lifted mainly due to the formation of quasi-1D 
orbital chains with Heisenberg-like dynamics. Concerning the separation 
of $T_{str}$ from $T_N$ in YVO$_3$, we think it is due to 
an increased tendency for the $xy$ orbital-selection by GdFeO$_3$-type 
distortions; this helps in fact the formation of $xz/yz$ doublets 
already above $T_N$. Below $T_{str}$, a short-range spin correlations and 
orbital dynamics is of quasi 1D nature, and $xz/yz$ doublet polarization 
is small. Essentially, the $xz/yz$ sector remains almost disordered for both 
quantum and entropy reasons~\cite{Kha01b,Ulr03,Sir03}. 
In our view, a complete classical order in the $xz/yz$ sector sets 
in only below a second structural transition at $T^{\star}_{str}\sim77$~K, 
determined by a competition between $\sim$1D spin-orbital physics 
and GdFeO$_3$-type distortions, which prefer a different ground 
state (more to come in Section 4). 

Apart from a neutron scattering experiments~\cite{Ulr03} challenging 
a classical JT picture for vanadates, we would like to quote here a recent 
paper~\cite{Yan04}, which observes that the vanadates become 
"transparent" for the thermal phonons {\it only} below a second 
transition at $T^{\star}_{str}$ (if present), in a so-called 
low-temperature $G$-phase, where we indeed expect that 
everything should be "normal" (described by a static orbital-angle physics). 
Enhanced phonon scattering on $xz/yz$ fluctuations, which suddenly 
disappears below $T^{\star}_{str}$, is very natural within 
the superexchange model. While it would be hard to understand this, 
if the orbital states both above and below $T^{\star}_{str}$ 
are classically ordered via the JT mechanism. Obviously, the thermal 
conductivity measurements are highly desired in titanites, in order to see 
as whether the phonon scattering remains unquenched down to low 
temperatures as expected from a superexchange picture, or will 
be like in manganites with a static orbitals. 
    
To summarize our present view on the role of orbital-lattice mechanism in 
perovskites: ({\it i}) The JT physics dominates in manganites, with a 
secondary role of SE interactions and extrinsic GdFeO$_3$-type 
distortions; ({\it ii}) In titanites and vanadates, the "orbital-angle" 
description is insufficient. It seems that orbital-lattice 
coupling plays a secondary, but {\it important} role providing a 
phase-selection out of a manifold of nearly degenerate many-body 
quantum states, offered by superexchange interactions. 
In vanadates with smaller A-site ions, though, a classical $G$ phase, 
favored by GdFeO$_3$-type distortions, takes over in the ground state, 
but the quantum spin-orbital states are restored again at finite 
temperature due to their larger spin-orbital entropy \cite{Kha01b,Ulr03}; 
--- a somewhat curious cooperation of quantum and thermal effects. 

\section{Lifting the orbital degeneracy by spin-orbital superexchange}
While the kinetic energy of electrons is represented in metals by 
the hopping $t$-Hamiltonian, it takes a form of spin-orbital superexchange 
in the Mott insulator. The superexchange interactions are obtained by 
eliminating the virtual doublon/holon states, a procedure which is 
justified as far as $t/U$ is small, and the resulting SE--scale 
$4t^2/U$ is smaller than $t$ itself. Near the Mott transition, 
a longer-range coupling and retardation effects, 
caused by a softening and unbinding of doublon/holon excitations are 
expected, and separation of the spin-orbital modes 
from an emergent low-energy charge and fermionic excitations 
becomes valid only locally. An example for the latter case is a cubic  
perovskite SrFeO$_3$, "bad" metal which is located just slightly below 
the Mott transition, or CaFeO$_3$ having in addition a weak charge-density 
modulation of low-energy holons/doublons. Here, both the superexchange 
interaction accounting for the high-energy charge fluctuations, 
{\it and} the low-energy/gapless charge excitations 
present near the transition, have to be considered on equal 
footing \cite{Kha06}. This picture leads to a strong competition 
between the superexchange and double-exchange processes, 
resulting in orbital disorder, a helical spin structure, and small-only 
Drude weight (quantifying the distance to the Mott-transition), as observed 
in ferrates \cite{Leb04}. 

We consider here conventional nearest-neighbor (NN) SE-models, 
assuming that the above criteria $4t^2/U<t$ is valid 
and the local spin-orbital degrees of freedom are protected 
by a charge gap. This is in fact consistent with 
spinwave measurements \cite{Kei00,Ulr02,Ulr03}, 
which can reasonably well be fitted by NN-exchange models 
in all compounds discussed in this section. 

In order to underline a difference between the spin exchange, 
described by a conventional Heisenberg interaction, and that in the orbital 
sector, we consider first the orbital-only models, and move then to the 
mechanisms which operate in the case of full spin-orbital Hamiltonians with 
different orbital symmetry and spin values.     

\subsection{Orbital-exchange, $e_g$ symmetry}

On the cubic lattice, the exchange of $e_g$ orbital quantum numbers 
is described by the Hamiltonian 
\begin{equation}
H_{orb} =\frac{2t^2}{U} \sum_{\langle ij\rangle}
\tau_i^{(\gamma)}\tau_j^{(\gamma)},
\label{ORB}
\end{equation}
where the pseudospin operators $2\tau^{(\gamma)}= 
\cos\phi^{(\gamma)}\sigma_i^{z}+\sin\phi^{(\gamma)}\sigma_i^{x}$~ 
have already been used in Eq.(\ref{Hn}). Here, the orientation of the 
bond $\langle ij \rangle$ is specified by the angle $\phi^{(\gamma)}=2\pi n/3$ 
with $n=1,2,3$ for $\gamma=a,b,c$ directions, respectively. 
(For the formal derivation of $H_{orb}$, consider a spin-polarized version 
of Eq.(\ref{Hn}) and set $\vec S_i \cdot \vec S_j=4$). 

As pseudospins in (\ref{ORB}) interact antiferromagnetically for all bonds, 
a staggered orbital ordered state is expected to be the ground state 
of the system. However, linear spin-wave theory, when applied to this  
Hamiltonian, leads to a gapless two-dimensional excitation 
spectrum\cite{Bri99}. This results in an apparent instability 
of the ordered state at any finite temperature, an outcome that sounds 
at least counterintuitive. Actually, the problem is even more severe: 
by close inspection of the orbiton-orbiton interaction corrections, 
we found that the orbiton self-energy diverges even at zero temperature, 
manifesting that the linear spin-wave expansion about a classical 
staggered orbital, N\'eel-like state {\it is not adequate} at all. 

The origin of these problems is as follows \cite{Kha01a}: 
By symmetry, there are only a finite number of directions, one of which 
will be selected as a principal axis for the quadrupole moment. 
Since this breaks only discrete symmetry, the excitations about
the ordered state must have a gap. 
A linear spin wave theory fails however to give the gap, because
Eq.(\ref{ORB}) acquires a rotational symmetry 
in the limit of classical orbitals. This results in an infinite 
degeneracy of classical states, and an {\it accidental} pseudo-Goldstone 
mode appears, which is however not a symmetry property of the original 
{\it quantum} model (\ref{ORB}). 
This artificial gapless mode leads to low-energy divergencies that arise
because the coupling constant for the interaction between orbitons does not 
vanish in the zero momentum limit, as it would happen for a true Goldstone 
mode. Hence the interaction effects are non-perturbative. 

At this point the order-from-disorder mechanism \cite{Tsv95} comes into
play: a particular classical state is selected so that the 
fluctuations about this state maximize the quantum energy gain, 
and a finite gap in the excitation spectra opens, because in the
ground state of the system the rotational invariance is broken.
To explore this point explicitly, we calculate 
quantum corrections to the ground state energy as a function of the angle
$\theta$ between $c$-axis and the moment direction. Assuming the latter is
perpendicular to $b$-axis, we rotate globally a pseudospin quantization axes as
$\sigma^z\rightarrow \sigma^z\cos\theta-\sigma^x\sin\theta$, 
$\sigma^x\rightarrow \sigma^x\cos\theta+\sigma^z\sin\theta$, 
and perform then orbital-wave expansion 
$\sigma_i^z=1-2a_i^{\dagger}a_i$, $\sigma_i^x\simeq a_i+a_i^{\dagger}$ 
around the classical N\'eel state, where the staggered moment is now 
oriented along the new $z$ direction. As a rotation of the quantization axes
changes the form of the Hamiltonian, one observes that the magnon excitation 
spectrum has an explicit $\theta$-dependence: 
\begin{equation}
\omega_{\bf p}(\theta) = 
2A\Bigl[1-\gamma_{\bf p}-\frac{1}{\sqrt{3}}\eta_{\bf p}\sin 2\theta-
\lambda_{\bf p}(1-\cos 2\theta)\Bigr]^{1/2}.  
\end{equation}
Here, $\gamma_{\bf p}=(c_x+c_y)/2$, $\eta_{\bf p}=(c_x-c_y)/2$, 
$\lambda_{\bf p}=(2c_z-c_x-c_y)/6$, and $c_{\alpha}=\cos p_{\alpha}$~; 
the energy scale $A=3t^2/2U$. Calculating the zero point 
magnon energy $E(\theta)=-\sum_{\bf p}(A-\frac{1}{2}\omega_{\bf p}(\theta))$, 
one obtains an effective potential for the staggered moment direction 
(three minima at $\theta=\phi^{(\gamma)}=2\pi n/3$), 
which at small $\theta$ reads as a harmonic one: 
$E(\theta) = const + \frac{1}{2}K_{eff}\theta^2$, 
with an effective ``spring'' constant $K_{eff} = A\kappa$. 
The parameter $\kappa$ is given by
\begin{equation}
\kappa = \frac{1}{3}\sum_{\bf p}\Bigl[\frac{2\gamma_{\bf p}}
{(1-\gamma_{\bf p})^{1/2}}
-\frac{\eta_{\bf p}^2}{(1-\gamma_{\bf p})^{3/2}}\Bigr]~ \approx 0.117~. 
\end{equation}

The physical meaning of the above calculation is that zero point
quantum fluctuations, generated by interactions between spin waves,
are enhanced when the staggered moment stays about a symmetric position
(one of three cubic axes), and this leads to the formation of 
the energy profile of cubic symmetry.
A breaking of this discrete symmetry results then in the magnon gap, which 
should be about $\sqrt{K_{eff}/M}$ in the harmonic approximation, 
where an ``effective inverse mass'' $1/M$ is of the order of the value of 
the magnon bandwidth $W=2\sqrt{2}A$. More quantitatively, the potential 
$E(\theta)$ can be associated with an effective uniaxial anisotropy term, 
$\frac{1}{2}K_{eff}\sum_{\langle ij\rangle_c}\sigma_i^z\sigma_j^z$,
generated by spinwave interactions in the symmetry broken phase. 
This low energy effective anisotropy leads to the gap
$\Delta = 2\sqrt{AK_{eff}}=2A\sqrt{\kappa} \sim 0.7A$, stabilizing 
a long-range ordering. The excitation gap compares 
to the full magnon bandwidth as $\Delta/W \simeq 0.24$. 
Almost the same gap/bandwidth ratio for the model (\ref{ORB}) was 
also obtained in Ref.~\cite{Kub02} by using a different method, 
{\it i.e.} the equation of motion. 

The above simple example illustrates a generic feature of orbital-exchange 
models: The interactions on different bonds are competing (like in the 
three-state Potts model), and, very different from the Heisenberg-like 
spin interactions, quantum effects are of crucial importance 
even for three-dimensional cubic system. In fact, the model (\ref{ORB}) 
has a finite classical gap and hence no much dynamics in 2D, 
thus fluctuations are {\it more important} in 3D. 
In this way, the orbital degeneracy provides a new root to 
frustrated quantum models in three dimensions, in addition to  
the conventional one driven by geometrical frustration. 

It should be interesting to consider the model (\ref{ORB}) 
for higher dimensional hypercubic lattices, letting the angle be 
$\phi^{(\gamma)}=2\pi n/d$, $n=1,...,d$. With increasing the number of 
bonds $\gamma=1,...,d$ the energy potential (having $d$-minima 
as function of the moment direction $\theta$) will gradually flatten, 
and hence the gap will eventually close in the limit of infinit dimensions. 
The ground state and excitations in that limit should be very peculiar. 

Finally, by considering a spin-paramagnetic case $\vec S_i\cdot\vec S_j=0$ 
in Eq.(\ref{Hn}), one would arrive again at the orbital model (\ref{ORB}), 
which leads to the $G$-type staggered order, different 
from what is seen in manganites well above T$_N$. Moreover, spin-bond 
fluctuations in the paramagnetic phase would wash out the three-minima 
potential, hence preventing the orbital order. This indicates again 
that the orbital transition at $\sim$800~K in LaMnO$_3$ is 
primarily not caused by the electronic superexchange. 

\subsection{Orbital-exchange, $t_{2g}$ symmetry: YTiO$_3$}

\begin{figure}
\centerline{\epsffile{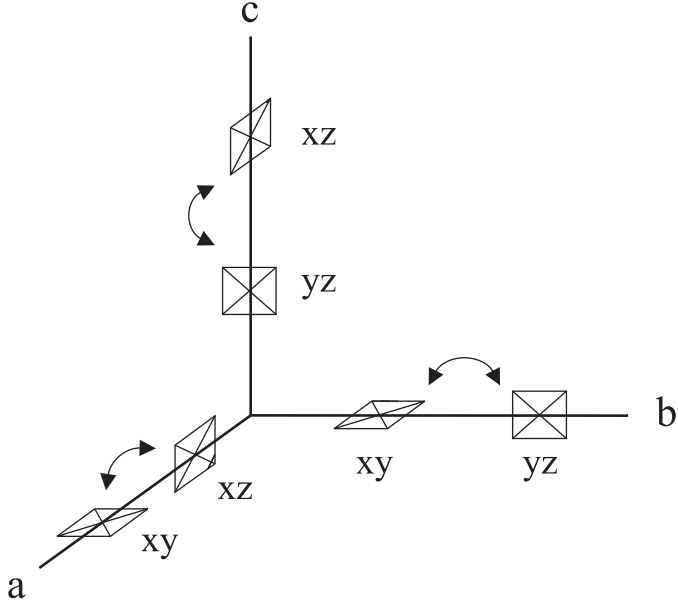}}
\vspace*{2ex}
\caption{For every bond of the cubic crystal, two out of three $t_{2g}$ 
orbitals are equally involved in the superexchange and may resonate.
The same two orbitals select also a particular component of angular
momentum. (After Ref.~\protect\cite{Kha01a}).
}
\label{fig1}
\end{figure}

Now, we discuss the $t_{2g}$-counterpart of the model (\ref{ORB}), 
which shows a more rich behavior. This is because of the higher, 
threefold degeneracy, and different hopping geometry of   
a planar-type $t_{2g}$ orbitals that brings new symmetry elements. 
The orbital order and fluctuations in $t_{2g}$ orbital-exchange 
model have been studied in Refs.\cite{Kha02,Kha03} in the context 
of ferromagnetic YTiO$_3$. The model reads (in two equivalent forms) as: 
\begin{eqnarray}
\label{ytio3tau}
H_{orb}&=&\frac{4t^2}{E_1}\sum_{\left\langle ij\right\rangle}
\big(\vec\tau_i\cdot \vec\tau_j+\frac{1}{4}n_i n_j\big)^{(\gamma)} \\
&=&\frac{2t^2}{E_1}\sum_{\left\langle ij\right\rangle}
(n_{i\alpha}n_{j\alpha}+n_{i\beta}n_{j\beta}
+\alpha_i^\dagger\beta_i\beta_j^\dagger\alpha_j 
+\beta_i^\dagger\alpha_i\alpha_j^\dagger\beta_j)^{(\gamma)}~. 
\label{ytio3}
\end{eqnarray}
This is obtained from Eq.(\ref{eq:original}) below by setting 
${\vec S}_i \cdot {\vec S}_j=1/4$ (and dropping a constant energy shift). 
The energy $E_1=U-3J_H$ corresponds to the high-spin virtual 
transition in the spin-polarized state of YTiO$_3$. In above equations,  
$\alpha\neq\beta$ specify the two orbitals active on a given bond direction 
$\gamma$ (see Fig.\ref{fig1}). For each $(\alpha\beta)^{(\gamma)}$-doublet, 
the Heisenberg-like pseudospin $\vec\tau^{(\gamma)}$:   
$\tau_z^{(\gamma)}=(n_{\alpha}-n_{\beta})/2$, 
$\tau_+^{(\gamma)}=\alpha^\dagger\beta$, 
$\tau_-^{(\gamma)}=\beta^\dagger\alpha$, 
and density 
$n^{(\gamma)}=n_{\alpha}+n_{\beta}$ are introduced. 

New symmetry elements mentioned above are: ({\it i}) a pseudospin 
interactions are of the Heisenberg form; thus, the orbital 
doublets like to form singlets just like spins do. 
({\it ii}) Apart from an obvious discrete cubic symmetry, the 
electron density on {\it each} orbital band is a conserved quantity. Formally, 
these conservation rules are reflected by a possibility of uniform 
phase transformation of orbiton operators, that is, 
$\alpha \rightarrow \alpha ~exp(i\phi_\alpha)$, which leaves the 
Hamiltonian invariant. Moreover, as $t_{2g}$-orbitals can hop only 
along two directions (say, $xy$-orbital motion is restricted to $ab$ 
planes), the orbital number is conserved on each plane separately. 

\begin{figure}
\epsfxsize=75mm
\centerline{\epsffile{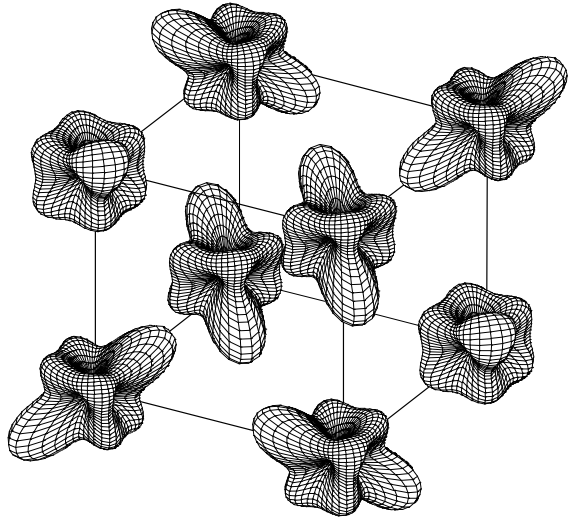}}
\caption{
$t_{2g}$-electron density in the quadrupole ordered state, 
calculated from Eq.~(\ref{eq:density}).
(After Ref.~\protect\cite{Kha02}).} 
\label{fig2}
\end{figure}

The above features make the ground state selection far more complicated 
than in case of $e_g-$orbitals, as it has in fact been indicated  
long ago~\cite{Kug75}. In short (see for the technical details 
Ref.\cite{Kha03}), the breaking of a discrete (cubic) symmetry is obtained 
via the order-from-disorder scenario again. It turns out, however, that in
this case quantum fluctuations select the body diagonals of the cube 
as a principal axes for the emerging quadrupole order parameter 
(see Fig.\ref{fig2}). The ordered pattern has a four-sublattice structure, 
and the density distribution for the first sublattice 
(with [111] as a principal axis) is described by a function: 
\begin{equation}
\rho_1 (\vec r) = 
\frac{1}{3} \bigl( d_{yz}^2 + d_{xz}^2 + d_{xy}^2 \bigr) 
+ \frac{2}{3} Q (d_{yz} d_{xz} + d_{yz} d_{xy} + d_{xz} d_{xy})~. 
\label{eq:density}
\end{equation} 
(Similar expressions can easily be obtained for other sublattices by 
a proper rotation of the quantization axes according to Fig.\ref{fig2}). 
Because of quantum fluctuations, the quadrupole moment $Q$, which controls 
the degree of orbital elongation, is anomalously small: $Q\simeq0.19$ 
(classically, $Q=1$). 

Surprisingly, not only quadrupole but also a magnetic ordering is equally good 
for the $t_{2g}$ orbital model. This corresponds to a condensation of 
complex orbitals giving a finite angular momentum, which is again small,  
$m_l\simeq 0.19\mu_B$. A magnetic pattern is of similar four-sublattice 
structure. Further, it turns out that quadrupole and magnetic orderings 
can in fact be continuously transformed to each other --- using a phase 
freedom present in the ground state --- and a mixed states do appear 
in between. We found that these phase degrees of freedom are the gapless 
Goldstone modes, reflecting the "orbital color" conservation rules 
discussed above. 

On the technical side, all these features are best captured by a radial gauge 
formalism, applied to the orbital problem in Ref.\cite{Kha03}. Within 
this approach, we represent the orbiton operators, entering in
Eq.(\ref{ytio3}), as $\alpha_i=\sqrt{\rho_{i\alpha}} e^{i \theta_{i\alpha}}$, 
thus separating a density and phase fluctuations. As a great advantage, 
this makes it possible to single out the amplitude fluctuations 
(of the short-range order parameters $Q$, $m_l$), responsible for discrete 
symmetry breaking, from a gapless phase-modes which take care of the 
conservation rules. This way, the ground state {\it condensate} wave-function 
was obtained as 
\begin{equation} 
\psi_{1,2,3,4}(\theta,\varphi)=\sqrt{\rho_0} 
\Bigl\{d_{yz} e^{i (\varphi + \theta)}
      \pm d_{zx} e^{i (\varphi - \theta)}
      \pm d_{xy} \Bigr\}.  
\label{condensate}
\end{equation} 
Here, $\rho_0\ll 1$ determines the amplitude of the local order parameter, 
while the phases $\varphi,\theta$ fix its physical nature --- whether it is of 
quadrupole or magnetic type. Specifically, quadrupole and magnetic 
orderings are obtained when $\varphi=\theta=0$, and 
$\varphi=\pi$, $\theta=\pi/3$, respectively. 
While short-range orbital ordering (a condensate fraction 
$\rho_0$) is well established at finite temperature via the order-from-disorder
mechanism, true long-range order (a phase-fixing) sets in at zero temperature
only. Slow space-time fluctuations of the phases $\varphi,\theta$ are
manifested in a 2D gapless nature of the orbital excitations. 

In Eq.(\ref{condensate}), we recognize the "orbital-angle" picture but, 
because the order parameteres are weak ($\rho_0=Q/3\sim0.06$), it represents 
only a small coherent fraction of the wave function; the main spectral 
weight of fluctuating orbitals is contained in a many-body wavefunction 
that cannot be represented in such a simple classical form at all. 
  
The {\it low-energy} behavior of the model is changed once perturbations 
from lattice distortions are included. An important effect is the deviation 
of the bond angles from $180^{\circ}$ of an ideal perovskite; this relaxes 
the orbital-color conservation rules, making a weak orbital order possible at
finite temperature. Physically, however, this temperature is confined to 
the spin ordering, since the interactions as shown in Eq.(\ref{ytio3}) 
are formed only after the spins get fully polarized, while fluctuating 
spins destroy the orbital order which is so fragile even in the ground state. 

A remarkable feature of the SE driven orbital order is that, although cubic
symmetry is locally lifted by a small quadrupole moment, the {\it bonds} 
remain perfectly the same, as evident from Fig.\ref{fig2}. This immediately 
explains a cubic symmetry of the spin-exchange couplings \cite {Ulr02}, and 
of the Raman light scattering from orbital fluctuations \cite {Ulr05}, --- 
the observations which seem so unusual within a crystal-field picture for 
YTiO$_3$. This indicates a dominant role of the superexchange 
mechanism in titanates.

{\it Orbital excitations in YTiO$_3$}.--- The superexchange 
theory predicts the following orbital excitation spectrum for 
YTiO$_3$ \cite{Kha03}:  
\begin{eqnarray} 
\omega_\pm ({\bf p})=W_{orb}
\bigl\{1-(1-2\varepsilon)(1-2f)(\gamma_1 \pm \kappa)^2 
 -2(\varepsilon -f)(\gamma_1 \pm \kappa)\bigr\}^{1/2}, 
\label{omegafinal}
\end{eqnarray} 
where subscript ($\pm$) indicates two orbiton branches,  
$\kappa^2=\gamma_2^2+\gamma_3^2$, and $\gamma_{1,2,3}$ are the 
momentum dependent form-factors $\gamma_1({\bf p})=(c_x+c_y+c_z)/3$, 
$\gamma_2({\bf p})=\sqrt3(c_y-c_x)/6$, $\gamma_3({\bf p})=(2c_z-c_x-c_y)/6$ 
with $c_{\alpha}=\cos p_{\alpha}$. Physically, the parameter 
$\varepsilon\simeq 0.2$ accounts for the many-body corrections 
stemming from the interactions between orbital waves, which 
stabilize a weak orbital order via the order-from-disorder mechanism. 
While the correction $f\sim0.1$ determines the orbital gap: it would be  
zero within the model (\ref{ytio3}) itself, but becomes finite once  
the orbital-nondiagonal hoppings (induced by octahedron tilting 
in GdFeO$_3$-type structure) are included in the calculations. 
Finally, the parameter $W_{orb}\simeq 2 (4t^2/E_1)$ represents an overall 
energy scale for the orbital fluctuations. By fitting the spin-wave data, 
$4t^2/E_1\simeq 60$~meV has been estimated for YTiO$_3$ in Ref.\cite{Kha03}; 
accordingly, $W_{orb}\sim 120$~meV follows. 

The energy scale $4t^2/E_1\simeq 60$~meV is in fact suggested also by a 
simple estimation, consider {\it e.g.} $t\sim 0.2$~eV and 
$E_1\leq 2.5$~eV inferred 
from optics \cite{Oki95}. However, it would be a good idea to "measure" 
this energy from optical experiments, as done for manganites~\cite{Kov04}. 
At low temperature, when a spin "filtering factor" 
$({\vec S}_i \cdot {\vec S}_j +3/4)$ is saturated for the high-spin 
transition --- see first line in Eq.(\ref{eq:original}) ---  
the ground state energy (per bond) is: 
\begin{equation}
\frac{4t^2}{E_1}
\Bigl[\big\langle\vec\tau_i\cdot \vec\tau_j+
\frac{1}{4}n_i n_j\big\rangle^{(\gamma)}
-\frac{1}{3}\Bigr]~=~-\frac{4t^2}{3E_1}(|E_0|+1)~, 
\end{equation}
consisting of a quantum energy $E_0\simeq -0.214$ (per site) calculated in 
Ref.\cite{Kha03}, and a constant stemming from a linear terms 
$n_i^{(\gamma)}$ in (\ref{eq:original}). Using now Eq.(\ref{sumrule}), 
we find that $K_1\simeq 0.81\times(4t^2/E_1)$, which can directly be
determined from the optical carrier number $N_{eff,1}$ once measured. 

{\it Orbital fluctuations in Raman light scattering}.--- The superexchange 
energy for orbital fluctuations, $W_{orb}\sim 120$~meV, is apparently 
consistent with the energy of a weak signal in the optical transmission data 
of Ref.\cite{Ruc05}. Namely, our interpretation, based on {\it two-orbiton} 
absorption (similar to the {\it two-magnon peak} in spin systems), gives 
$2W_{orb}\sim 240$~meV for the peak position (without a phonon assisting 
the process), as observed. The same characteristic energy follows also 
for the Raman scattering \cite{Ulr05}, which is derived from a two-orbiton 
propagator with proper matrix elements. However, the most crucial
test is provided by a symmetry of the orbital states, controlling 
the polarization dependences. The superexchange theory outlined above 
predicts a {\it broad} Raman band (due to a many-body interactions between 
the orbital excitations), with the 
polarization rules of cubic symmetry. Our superexchange-Raman theory is 
conceptually identical to that for the light scattering on spin fluctuations, 
and, in fact, the observed orbital-Raman lineshapes are very similar 
to those of spin-Raman response in cuprates. On the contrary, we found, 
following the calculations of Ref.\cite{Ish04}, that the polarization 
rules for the {\it lattice driven} orbital states (\ref{function}) 
in YTiO$_3$ strongly disagree with cubic symmetry: 
The energy positions are completely 
different for the $c$ axis and $ab$ plane polarizations. Such a strong 
anisotropy is imposed by a "broken" symmetry of the lattice: 
in a crystal-field picture, the orbital state is "designed" to fit these 
distortions (by tuning the orbital-angles). 

Comparing the above two models, (\ref{ORB}) and (\ref{ytio3tau}), we see 
completely different low-energy behavior --- while a finite temperature 
long-range order is protected by a gap in former case, no gap 
is obtained for the $t_{2g}$ orbitals. Physically, this is related to 
the fact that $t_{2g}$ triplet accommodates not only the electric 
quadrupole moment but --- different from $e_g$ doublet --- also a true 
magnetic moment which is a vector like spin. 
It is this dual nature of the $t_{2g}$ triplet --- a Potts-like 
quadrupole {\it and} Heisenberg-like vector ---  which lies 
at the origin of rich physics of the model (\ref{ytio3tau}). 

\subsection{$e_g$ spin-orbital model, spin one-half} 
Let us move now to the spin-orbital models that describe a simultaneous 
exchange of both spin and orbital quantum numbers of electrons. We 
start with the SE model for $e_g$ holes as in perovskites like KCuF$_3$, 
neglecting the Hund's rule corrections for simplicity. 
On a three-dimensional cubic lattice it takes the form~\cite{Kug82}: 
\begin{eqnarray}
\label{HAM2}
H=J\sum_{\langle ij\rangle}\big(\vec S_i \cdot \vec S_j+\frac{1}{4}\big) 
\big(\frac{1}{2}-\tau_i\big)^{(\gamma)}
\big(\frac{1}{2}-\tau_j\big)^{(\gamma)},  
\end{eqnarray}
where $J=4t^2/U$, and $\tau^{(\gamma)}$ are the $e_g$-pseudospins defined 
above. The main feature of this model --- suggested by the 
very form of Hamiltonian (\ref{HAM2}) --- is the strong interplay between 
spin and orbital degrees of freedom. It was first recognized 
in Ref.~\cite{Fei97}, that this simple model contains rather nontrivial 
physics: the classical N\'eel state is infinitely degenerate in the orbital 
sector thus frustrating orbital order and {\it vice versa}; this extra 
degeneracy must be lifted by some mechanism (identified later-on in 
Ref.\cite{Kha97}).  

We first notice that the effective spin-exchange constant in 
this model is definite positive for any configuration of orbitals 
(as $\tau\leq1/2$), where its value can vary from 
zero to $J$, depending on the orientation of orbital pseudospins. 
We therefore expect a simple two-sublattice antiferromagnetic, G-type, 
spin order. There is however a problem: a classical G-type ordering has
cubic symmetry and can therefore not lift the orbital degeneracy. 
In more formal terms, the spin part 
$({\vec S}_i \cdot {\vec S}_j +1/4)$ of the Hamiltonian (\ref{HAM2}) 
simply becomes zero in this state for all bonds, so that these orbitals 
effectively do not interact --- they are completely uncorrelated. 
In other words, we gain no energy from the orbital interactions. 
This shows that from the point of view of the orbitals the classical 
N\'eel state is energetically a very poor. 

The mechanism for developing intersite orbital correlations 
(and hence to gain energy from orbital ordering) must involve a strong 
deviation in the spin configuration from the N\'eel state --- a deviation 
from $\langle \vec{S}_i \cdot \vec{S}_j \rangle = -\frac{1}{4}$. 
This implies an intrinsic tendency of the system to develop low-dimensional
spin fluctuations which can most effectively be realized by an ordering 
of elongated $3z^2-r^2$ orbitals [that is, $\alpha_i=0$ in Eq.(\ref{1angle})]. 
In this situation the effective spin interaction 
is {\it quasi one-dimensional}, so that spin 
fluctuations are enhanced as much as possible and large quantum energy 
is gained from the bonds along the $3z^2-r^2$ orbital chains. Since 
$\langle \vec{S}_i \cdot \vec{S}_j+\frac{1}{4}\rangle_{c}<0$, 
the effective orbital-exchange constant that follows from (\ref{HAM2}) 
is indeed ferromagnetic, thus supporting $3z^2-r^2$ type uniform order. 
At the same time the cubic symmetry is explicitely broken, 
as fluctuations of spin bonds are different in different directions.
This leads to a finite splitting of $e_g-$levels, and therefore an 
orbital gap is generated. One can say that in order to stabilize 
the ground state, a quadrupole order and anisotropic spin fluctuations support 
and enhance each other --- one recognizes here the order-from-disorder 
phenomenon again. 

More quantitatively, the expectation value of the spin-exchange coupling 
along the $c$-axis, given by strong $3z^2-r^2$ orbital overlap 
[consider $\tau^{(c)}=-1/2$ in Eq.(\ref{HAM2})], is $J_c=J$, while 
it is only small in the $ab$-plane: $J_{ab}=J/16$. Exchange energy 
is mainly accumulated in $c$-chains and can be approximated as
$J_{c} \langle \vec S_i\cdot \vec S_j+\frac{1}{4}\rangle_{c} +
2J_{ab}\langle \vec S_i\cdot \vec S_j+\frac{1}{4}\rangle_{ab}\simeq -0.16J$ 
per site (using $\langle\vec S_i\cdot\vec S_j\rangle_c=1/4-\ln 2$ for 
1D and assuming $\langle\vec S_i\cdot\vec S_j\rangle_{ab}\sim 0$). 
On the other hand, $x^2-y^2$ orbital ordering results in the two-dimensional 
magnetic structure ($J_{a,b}=9J/16, J_c=0 $) with a much smaller 
energy gain $\simeq -0.09J$. 

From the technical point of view, it is obvious that a conventional 
expansion around the classical N\'eel state would fail to remove the 
orbital degeneracy: Only quantum spin fluctuations can lead to orbital 
correlations. This is precisely the reason why one does not obtain 
an orbital gap in a linear spin-wave approximation, and low-energy 
singularities appear in the calculations \cite{Fei97}, leading to an 
{\it apparent} collapse of the spin and orbital order. 
However, as demonstrated in Ref.\cite{Kha97}, the long-range 
orderings are {\it stable} against fluctuations, and no 
"quantum melting" does in fact occur. The singularities 
vanish once quantum spin fluctuations are explicitely taken into 
account in the calculations of the orbiton spectrum. These fluctuations 
generate a finite gap (of the order of $J/4$) for a single orbital as well 
as for any composite spin-orbital excitation. The orbital gap removes 
the low-energy divergencies, protecting the long-range spin order. 
However, spin order parameter is strongly 
reduced to $\langle S^z \rangle\simeq 0.2$, due to the 
quasi-one dimensional structure of spin-orbital correlations. 
Such a spin reduction is generic to spin-orbital models and occurs also 
in the $t_{2g}$ case, but the mechanism is quite different, as we see below. 

Physically, because of the strong spatial anisotropy of the $e_g$-orbitals, 
it is impossible to optimize the interactions in all the bonds 
simultaneously; this results in orbital frustration. The frustration is 
removed here by reducing the effective dimensionality of the interactions, 
specifying strong and weak bonds in the lattice. (We may speak of the  
"Peierls effect" without the phonons; this is essentially what happens 
in vanadates, too, see later). At the same time tunneling between different 
orbital configurations is suppressed: the spin fluctuations produce 
an energy gap for the rotation of orbitals. A similar mechanism of 
resolving the frustrations by using the orbital degrees of freedom 
has recently been discussed in Ref.~\cite{Tsu03} for vanadium spinels. 

Our last point concerns the temperature scales, $T_{orb}$ and $T_N$. They are 
in fact {\it both} controlled by the same energy, that is the orbital gap 
$\Delta\sim J/4$\cite{Kha97}. Once the quadrupole order is lost at 
$T_{orb}\sim\Delta$ due to the flat 2D orbital modes, spin order will also 
collapse. Alternatively, thermal destruction of the spin correlations 
washes out the orbital gap. 
Thus, $T_{orb}\sim T_N\sim\Delta$ in the $e_g$ exchange-model alone. 
[To obtain $T_{orb}\gg T_N$ as commonly observed in $e_g$ compounds 
experimentally, the orbital frustration should be eliminated 
by lattice distortions]. 
In $t_{2g}$ systems, however, a "delay" of $T_{orb}$, 
and even $T_{orb}\ll T_N$, {\it is possible}. This is because the 
$t_{2g}$ orbitals are far more frustrated than the Heisenberg spins are. 
Such an extreme case is in order to be analyzed now. 

\subsection{$t_{2g}$ spin-orbital model, spin one-half: LaTiO$_3$} 
We consider first a full structure of the SE Hamiltonian in titanates. 
Virtual charge fluctuation spectra for the Ti-Ti pair is represented 
by a high-spin transition at $E_1=U-3J_H$ and low-spin ones at energies 
$E_2=U-J_H$ and $E_3=U+2J_H$. Here, $U=A+4B+3C$ and $J_H=3B+C$ are 
the intraorbital repulsion and Hund's coupling in the Ti$^{2+}$ 
excited state, respectively\cite{Gri61}. From the optical data of 
Ref.\cite{Oki95}, one may infer that these transitions are located  
within the energy range from $\sim 1$~eV to $\sim 4$~eV. Because of 
the small spin value, the Hund's splittings are apparently less than 
the linewidth, thus these transitions strongly overlap in optics. 
Indeed, {\it the free-ion} value $J_H\simeq 0.59$~eV \cite{Gri61} for 
Ti$^{2+}$ gives $E_2-E_1 \simeq 1.2$~eV, compared 
with a $t_{2g}$ bandwidth $\sim 2$~eV. (We believe that $J_H$ is further 
screened in the crystal, just like in manganites\cite{Kov04}). 
Experimentally, the temperature dependence 
of the optical absorption may help to resolve the transition energies, 
and to fix thereby the values of $U$ and $J_H$ in crystal. For YTiO$_3$, 
we expect that the $E_1-$band should increase (at the expense of the 
low-spin ones) as the sample is cooled down developing ferromagnetic 
correlations. The situation in AF LaTiO$_3$ is, however, much more delicate 
because of strong quantum nature of spins in this material (recall that 
the spin-order parameter is anomalously small), and because of the absence 
of {\it a cooperative} orbital phase transition. Thus, we expect 
no sizable thermal effects on the spectral weight distribution within 
the $d_id_j-$optical multiplet in LaTiO$_3$. (Optical response theory 
for LaTiO$_3$, where the quantum effects are of vital importance, is still 
lacking). This is in sharp contrast to manganites, where the classical 
spin- and orbital-orderings lead to a dramatic spectral weight 
transfer at $T_N$ and $T_{str}$ \cite{Kov04}. 

The above charge fluctuations lead to the SE Hamiltonian~\cite{Kha01a,Kha03} 
which we represent in the following form: 
\begin{eqnarray} 
\label{eq:original} 
H&=&\frac{2t^2}{E_1}\Bigl({\vec S}_i\cdot{\vec S}_j+\frac{3}{4}\Bigl) 
\Bigl(A_{ij}^{(\gamma)}-\frac{1}{2}n_i^{(\gamma)}-
\frac{1}{2}n_j^{(\gamma)}\Bigl)                    \\
 &+&\frac{2t^2}{E_2}\Bigl({\vec S}_i\cdot{\vec S}_j-\frac{1}{4}\Bigl) 
\Bigl(A_{ij}^{(\gamma)}+\frac{1}{2}n_i^{(\gamma)}+
\frac{1}{2}n_j^{(\gamma)}\Bigl)           \nonumber \\ 
 &+&\Bigl(\frac{2t^2}{E_3}-\frac{2t^2}{E_2}\Bigl) 
\Bigl({\vec S}_i\cdot{\vec S}_j-\frac{1}{4}\Bigl)
\frac{2}{3}B_{ij}^{(\gamma)}~.            \nonumber
\end{eqnarray} 
The spin-exchange constants (which determine the magnon spectra) 
are given by a quantum-mechanical average of the following operator:   
\begin{eqnarray} 
\hat J_{ij}^{(\gamma)}=J\Bigl[\frac{1}{2}(r_1+r_2)A_{ij}^{(\gamma)}  
-\frac{1}{3}(r_2-r_3)B_{ij}^{(\gamma)} 
-\frac{1}{4}(r_1-r_2)(n_i+n_j)^{(\gamma)}\Bigr],  
\label{Jgamma} 
\end{eqnarray} 
where $J=4t^2/U$. The parameters $r_n=U/E_n$ take care of the $J_H$-multiplet 
splitting, and $r_n=1$ in the limit of $J_H=0$. One should note that the 
spin-exchange constant is {\it only a fraction} of the full energy scale, 
represented by $J$, because of the compensation between contributions 
of different charge excitations $E_n$. This is typical when orbital 
degeneracy is present, but more pronounced for $t_{2g}$ systems where the 
spin interaction may have either sign even in the $J_H=0$ limit, see below.  
   
The orbital operators $A_{ij}^{(\gamma)}$, $B_{ij}^{(\gamma)}$  
and $n_i^{(\gamma)}$ depend on the bond direction $\gamma$, and 
can be represented in terms of constrained particles  
$a_i$, $b_i$, $c_i$ with $n_{ia}+n_{ib}+n_{ic}=1$  
corresponding to $t_{2g}$ levels of $yz$, $zx$, $xy$ symmetry, respectively. 
Namely, 
\begin{eqnarray} 
A_{ij}^{(c)}&=&n_{ia}n_{ja}+n_{ib}n_{jb}
+a_i^\dagger b_i b_j^\dagger a_j 
+b_i^\dagger a_i a_j^\dagger b_j, 
\label{eq:A_ab} 
\\ 
B_{ij}^{(c)} & = & n_{ia}n_{ja}+n_{ib}n_{jb} 
+ a_i^\dagger b_i a_j^\dagger b_j 
+ b_i^\dagger a_i b_j^\dagger a_j, 
\nonumber 
\label{eq:B_ab}  
\end{eqnarray} 
and $n_i^{(c)}=n_{ia}+n_{ib}$, for the pair along the $c$ axis. 
Similar expressions are obtained for the $a$ and $b$ bonds, 
by replacing $(ab)-$doublets by $(bc)$ and $(ca)$ pairs, respectively. 
It is also useful to represent $A_{ij}^{(\gamma)}$ and $B_{ij}^{(\gamma)}$ 
in terms of pseudospins:  
\begin{eqnarray} 
\label{eq:A_tau} 
A_{ij}^{(\gamma)}  
=2\big(\vec\tau_i\cdot\vec\tau_j+\frac{1}{4}n_i n_j\big)^{(\gamma)}, 
\;\;\;\;\; 
B_{ij}^{(\gamma)}  
=2\big(\vec \tau_i \otimes \vec \tau_j+\frac{1}{4}n_i n_j\big)^{(\gamma)},  
\end{eqnarray} 
where ${\vec \tau}_i^{(\gamma)}$ operates 
on the subspace of the orbital doublet $(\alpha,\beta)^{(\gamma)}$  
active on a given $\gamma$-bond (as already explained above), 
while a symbol $\otimes$ denotes a product $\vec\tau_i\otimes\vec\tau_j=
\tau_i^z\tau_j^z+(\tau_i^+\tau_j^+ + \tau_i^-\tau_j^-)/2$.

At large $J_H$, the ground state of the Hamiltonian (\ref{eq:original}) 
is obviously ferromagnetic, imposed by the largest $E_1-$term, and the problem
reduces to the model (\ref{ytio3tau}), in which:  
({\it i}) the orbital wave function is described by Eq.(\ref{condensate}) but, 
we recall that this is only a small condensate fraction; 
({\it ii}) low-energy excitations are the 2D, gapless, two-branch, 
Goldstone phase-modes. Concerning the spin excitations, we may 
anticipate some nontrivial things even in a ferromagnetic state. 
Once a magnon is created, it will couple to the orbital phase 
modes and {\it vice versa}. At very large $J_H$, this coupling is 
most probably of perturbative character but, as $J_H$ is decreased, 
a bound states should form between the spin and orbital Goldstone modes. 
This is because the magnons get softer due to increased contributions 
of the $E_2$ and $E_3$ terms. Evolution of the excitation spectra, 
and the nature of quantum phase transition(s) with decreasing 
$J_H$ [at which critical value(s)? of which order?] have not 
yet been addressed so far at all. Needless to say, 
the finite temperature behavior should be also nontrivial 
because of the 2D modes --- this view is supported also by Ref.\cite{Har03}.  

Looking at the problem from the other endpoint, $J_H=0, E_n=U$, where 
the ferromagnetic state is certainly lost, one encounters the 
following Hamiltonian: 
\begin{equation}
H=2J\sum_{\left\langle ij \right\rangle}
\big({\vec S}_i\cdot{\vec S}_j+\frac{1}{4}\big) 
\big(\vec \tau_i\cdot\vec \tau_j+\frac{1}{4}n_i n_j\big)^{(\gamma)}.
\label{Heta0}
\end{equation}
(An unessential energy shift, equal to $-J$, is not shown here). This model  
best illustrates the complexity of $t_{2g}$ orbital physics in perovskites. 
Its orbital sector, even taken alone as in (\ref{ytio3tau}), is nearly 
disordered; now, a fluctuating spin bonds will introduce strong disorder 
in the orbital sector, "deadly" affecting the orbital-phase modes, and 
hence the long-range coherence which was already so weak. Vice versa, 
the orbital fluctuations do a similar job in the spin sector; --- thus, 
the bound states mentioned above and spin-orbital entanglement 
are at work in full strength now. 

A while ago \cite{Kha00}, we proposed that, in the ground state, the 
model (\ref{Heta0}): ({\it i}) has a weak spin order of $G$-type which 
respects a cubic symmetry; ({\it ii}) the orbitals are fully disordered. 
Calculations within the framework of $1/N-$expansion, 
supporting this proposal, have been presented in that work. 
Here, we would like to elaborate more on physical ideas 
that have led to the orbital-liquid picture. 

Obviously, quantum dynamics is crucial to lift a macroscopic degeneracy 
of classical states in the model (\ref{Heta0}), stemming from an infinite 
number of the "orbital-color conservation" rules discussed above. 
Various classical orbital patterns (like a uniform $(xy+yz+zx)/\sqrt 3$, 
$xy$ orderings, {\it etc.}) leave us with Heisenberg spins alone, 
and hence give almost no energy gain and are ruled out. 
Quasi-1D orbital order like in the case of the $e_g$ model (\ref{HAM2}) is 
impossible because of a planar geometry of the $t_{2g}$ orbitals. 
Yet, the idea of a (dynamical) lowering of the effective dimensionality 
is at work here again, but underlying mechanism is radically different 
from that in $e_g$ case.  

The key point is a possibility to form orbital singlets. Consider, say, 
the exchange pair along the $c$ direction. {\it If} both ions are occupied 
by active orbitals ($n_i^{(c)}=n_j^{(c)}=1$), one obtains the interaction of 
the form $2J({\vec S}_i\cdot{\vec S}_j+1/4)(\vec\tau_i\cdot\vec\tau_j+1/4)$ 
that shows perfect symmetry between spin and orbital pseudospin. 
The pair has sixfold degeneracy in the lowest energy state: both 
{\it spin-triplet$\otimes$orbital-singlet} and 
{\it spin-singlet$\otimes$orbital-triplet} 
states gain the same exchange energy equal to $-J/2$. In other words, 
spin exchange constant may have equally strong ferromagnetic and 
antiferromagnetic nature depending on the symmetry of the orbital
wavefunction. This violates the classical Goodenough-Kanamori rules, in which 
ferromagnetic spin exchange occurs only at finite Hund's coupling and 
hence is smaller by factor of $J_H/U$. In this respect, $t_{2g}$ superexchange 
clearly differs from the $e_g$ model (\ref{HAM2}), in which the spin-exchange 
interaction is positively definite because no orbital singlets can be formed 
in that case. 

When such $t_{2g}$ pairs form 1D chain, one obtains a model which has
been investigated under the name {\it SU(4)} model~\cite{Li98,Fri99}. 
A large amount of quantum energy ($-0.41J$ per site) is gained in this 
model due to resonance between the local configurations 
{\it spin-triplet$\otimes$orbital-singlet} and 
{\it spin-singlet$\otimes$orbital-triplet}. As a result of this resonance, 
low-energy excitations are of composite spin-orbital nature. 
In a cubic lattice, the situation is more complicated, as {\it SU(4)} 
spin-orbital resonance along one direction necessarily frustrates 
interactions in the remaining two directions which require different 
orbital pairs (see Fig.\ref{fig1}). Given that {\it SU(4)} chain physics 
so ideally captures the correlations, one is nevertheless tempted to 
consider a "trial" state: the $xy$ orbital is empty, while $xz/yz$ doublets 
(together with spins) form {\it SU(4)} chains along the $c$ axis --- 
a kind of spin-orbital nematic, with a pronounced directionality of the
correlations. Accounting for the energy-lost on 
a "discriminated" (classical) $ab$ plane bonds 
on a mean-field level ($J/8$ per site), we obtain $E_0=-0.29J$ 
for our trial state, which is by far better than any static 
orbital state, and also better than the ferromagnetic state with fluctuating
orbitals ($E_0=-0.214J$ \cite{Kha03}). Once the $xy$ orbital is suppressed 
in our trial state, the interchain couplings read as 
\begin{equation}
H_{ij}^{(a/b)}=J\big({\vec S}_i\cdot{\vec S}_j+\frac{1}{4}\big) 
\big(\frac{1}{2}\pm\tau_i^z\big)\big(\frac{1}{2}\pm\tau_j^z\big), 
\label{Hab}
\end{equation}  
where $\pm$ sign refers to $a/b$ bond directions. In the ground state, 
these couplings may induce a weak ordering (staggered between 
the chains) in both sectors, which, however, should not affect much 
the intrachain {\it SU(4)} physics, by analogy with $\sim$1D spin 
systems~\cite{Sch96}. (This should be an interesting point to consider). 

An assumption made is that a quadrupole order parameter, $Q=n_a+n_b-2n_c$, 
responsible for the chain structure, is stabilized by order-from-disorder 
as in case of $e_g$ quadrupoles in (\ref{HAM2}), or as it happens 
in the {\it spin-one} model for vanadates (see later). However, the $t_{2g}$ 
quadrupole is highly quantum object, as wee have seen above in the context 
of YTiO$_3$, and it is hard to imagine that the above structure will survive 
against the $xy$ orbital intervention, that is, "cutting" the {\it SU(4)} 
chains in small pieces and their orientational disorder. One may 
therefore think of a liquid of "{\it SU(4)}--quadruplets" (a minimal building 
block to form a spin-orbital singlet\cite{Li98,Fri99}). This way, one arrives
at an intuitive picture of dynamical patterns where the local physics is 
governed by short-range {\it SU(4)} correlations, like in quantum dimer
models. As a crude attempt to capture the local {\it SU(4)} physics, 
we applied the $1/N-$expansion to the model (\ref{Heta0}), introducing a
bond-operator of mixed spin-orbital nature. A quadrupole disordered 
state was found indeed lower in energy ($E_0\simeq -0.33J$)\cite{Kha00} 
than a nematic state just discussed. As a $1/N-$expansion usually 
underestimates the correlations, we think that a quadrupole disordered 
state best optimizes the overall quantum energy of the 
$t_{2g}$ spin-orbital superexchange. An additional energy profit 
is due to the involvement of all three orbital flavors --- a natural 
way of improving a quantum energy gain. The nature of orbital excitations
is the most fundamental problem. Tentatively, we believe that a 
pseudogap must be present protecting a liquid state; this has already  
been indicated in Ref.\cite{Kha00} (as an orbital gap, stemming formally from 
the pairing effects within $1/N-$expansion). 

An important point is that spins and orbitals in the 3D model
~(\ref{Heta0}) are not equivalent. In the spin sector, the Heisenberg 
interactions on different bonds cooperate supporting the spin long-range 
order (albeit very weak) in the ground state. It is the orbital 
frustration which brings about an unusual quantum physics in a 3D system. 
When orbitals are disordered, the expectation value of the spin-exchange
constant, given by Eq.(\ref{Jgamma}), is of AF sign at small $J_H$, supporting 
a weak spin-$G$ order in the ground state, on top of underlying quantum
spin-orbital fluctuations. Important to note is that the local {\it SU(4)} 
physics is well compatible with a weak spin staggering. The main 
ingredient of the theory of Ref.\cite{Kha00} is a local {\it SU(4)} 
resonance, which operates on the scale of $J$ and 
lifts the orbital degeneracy without symmetry breaking. A remote analogy 
can be drawn with a dynamical JT physics: --- the role of phonons are
played here by spin fluctuations, and an entangled {\it SU(4)} spin-orbital 
motion is a kind of vibronic state but living on the bonds. 
While orbital-lattice vibronic states are suppressed by classical structural 
transitions, the orbital-spin {\it SU(4)} resonance may survive 
in a lattice due to quantum nature of spins one-half and orbital frustration,  
and may lead to the orbital disorder in the 3D lattice --- this is the 
underlying idea. 

A weak staggering of spins (while the orbitals are disordered) is due to the 
spin-orbital asymmetry for the 3D lattice. 
The Hund's coupling $J_H$ brings about yet another asymmetry 
between the two sectors, but this is now in favor of spin
ferromagnetism. $J_H$ changes a balance between two different (AF and F) spin 
correlations within a {\it SU(4)} resonance, and, eventually, a ferromagnetic 
state with a weak 3D quadrupole order (Fig.\ref{fig2}) is stabilized. Are 
there any other phases in between? Our {\it tentative} answer is "yes", and 
the best candidate is the spin-orbital nematic discussed above. This state
enjoys a fully developed {\it SU(4)} physics along the $c$ direction,
supported by orbital quadrupole ordering ($xy$ orbital-selection). 
The $xy$ orbital gap, induced in
such a state by $J_H$ in collaboration with order-from-disorder effect, 
is still to be quantified theoretically. In this intermediate phase, 
spin and $xz/yz$ doublet correlations are both AF 
within the planes (see Eq.\ref{Hab}), but different along the {\it SU(4)} 
chains: more ferro (than AF) for spins, and the other way round in the orbital
sector. Thus, we predict an intermediate phase with a {\it weak} 
spin-$C$ and orbital-$G$ order parameters. Our 
overall picture is that of the {\it three competing phases}: 
(I) spin-ferro and orbitals as in Fig.\ref{fig2}, stable at large $J_H$; 
(II) spin-$C$, doublets $xz/yz$ are staggered, $xy$ occupation 
is less than 1/3; (III) spin-$G$/orbital-liquid at small $J_H$. 
From the experience in vanadates (see next section), we suspect that 
a tight competition between these states may occur for realistic 
$J_H$ values. The first (last) states are the candidates 
for YTiO$_3$ (LaTiO$_3$); it should be a good idea looking for the 
intermediate one at compositions or compounds "in between". 
Needless to say, that all these three states are 
highly anomalous (compare with 3D Heisenberg systems), because the classical 
orderings here are just a secondary effects on top of underlying {\it SU(4)} 
quantum fluctuations (or of pure orbital ones at large $J_H$). 

Physically, $J_H-$tuning is difficult but can be somewhat mimicked by 
a variation of the Ti-O-Ti bond angle $\theta$ ({\it e.g.}, by pressure). 
A deviation of it from 180\% gives an additional term in the spin-exchange 
through the small $t_{2g}-e_g$ overlap as it has been pointed out 
in Ref.\cite{Moc01}. According to Ref.\cite{Kha03}, 
this term supports ferromagnetism 
{\it equally in all three} directions (different from Ref.\cite{Moc01}). 
Thus, such a term: $-J'\vec S_i\cdot\vec S_j$ with $J'\propto\sin^2\theta$~ 
\cite{Kha03} does not break a cubic symmetry itself, and hence may 
perfectly drive the above phase transitions. A pronounced quantum nature 
of the competing phases (because of {\it quantum orbitals}) may lead to 
a "soft" character of transitions, as suggested in Ref.\cite{Kha03}. 
Yet another explanation, based on {\it classical orbital} description, 
has been proposed in Ref.\cite{Moc01}, predicting 
the spin-$A$ phase as an intermediate state. Thus, the predictions 
of a quantum and classical orbital pictures are very 
different: the spin-$C$ {\it versus} the spin-$A$ type 
intermediate state, respectively. This offers a nice 
opportunity to discriminate between the electronic and 
lattice mechanisms of lifting the orbital degeneracy in titanates.      
    
Summarizing, the Hamiltonians (\ref{eq:original}) and (\ref{Heta0}) 
are the big "puzzles", providing a very interesting playground for theory. 
In particular, the phase transitions driven by $J_H$ are very intriguing. 
Concerning again the relation to the titanites: 
While the most delicate and interesting low-energy problems 
are (unfortunately) eliminated by weak perturbations like 
lattice distortions, the major physics --- a local {\it SU(4)} 
resonance --- should be still intact in LaTiO$_3$. This 
view provides {\it a hitherto unique} explanation for: 
({\it i}) an anomalous spin reduction (due to a quantum magnons 
involved in the spin-orbital resonance, see Ref.\cite{Kha00}); 
({\it ii}) the absence of a cooperative structural transition (the orbital 
liquid has no degeneracy, hence no JT instability at small coupling); 
({\it iii}) nearly ideal cubic symmetry of the spin and Raman responses 
in both LaTiO$_3$ and YTiO$_3$ (in full accord with our theory). The 
identification of the predicted intermediate spin-$C$ phase is a challenge 
for future experiments. 

\subsection{$t_{2g}$ spin-orbital model, spin one: LaVO$_3$} 
In the model for titanites, a quantum nature of spins one-half was 
essential; to make this point more explicit, we consider now similar 
model but with the higher spin, $S=1$. Apart from its direct 
relevance to pseudocubic vanadates AVO$_3$, the model provides 
yet another interesting mechanism of lifting the degeneracy by 
SE interactions: here, the formation of the quantum orbital chains 
is the best solution \cite{Kha01b}. 

The interactions between $S=1$ spins of V$^{3+}$ ions arise from 
the virtual excitations $d^2_id^2_j\rightarrow d^1_id^3_j$, and 
the hopping $t$ is allowed only between two out of three $t_{2g}$ orbitals, 
just as in titanites. The $d^3_i$ excited state may be either 
({\it i}) a high-spin $^4A_2$ state, or one of a low-spin ones: 
({\it ii}) the degenerate $^2E$ and $^2T_1$ states, or 
({\it iii}) a $^2T_2$ level. The excitation energies are $E_1=U-3J_H$, 
$E_2=U$ and $E_3=U+2J_H$, respectively \cite{Gri61}, 
where $U=A+4B+3C$ and $J_H=3B+C$. For the free ion V$^{2+}$, 
one has $J_H\simeq 0.64$~eV \cite{Gri61} but this should be screened 
in crystal to $\simeq 0.5$~eV as suggested in Ref.\cite{Kha04a}. 
Correspondingly, the SE Hamiltonian consists of three contributions, 
like in Eq.~(\ref{eq:original}), but a different form as obtained 
in Ref.\cite{Kha01b} is more instructive here: 
\begin{equation}
H=\sum_{\langle ij\rangle}\left[({\vec S}_i\cdot{\vec S}_j+1)
  {\hat J}_{ij}^{(\gamma)}+{\hat K}_{ij}^{(\gamma)}\right].
\label{model}
\end{equation}
In terms of operators $A_{ij}^{(\gamma)}$, $B_{ij}^{(\gamma)}$ 
and $n_i^{(\gamma)}$ introduced previously in 
Eqs.(\ref{eq:A_ab})--(\ref{eq:A_tau}), the orbital operators 
${\hat J}_{ij}^{(\gamma)}$ and ${\hat K}_{ij}^{(\gamma)}$ read as follows: 
\begin{eqnarray}
\label{orbj}
{\hat J}_{ij}^{(\gamma)}&=&\frac{J}{4}\left[(1+2\eta R)A_{ij}^{(\gamma)}
-\eta r B_{ij}^{(\gamma)}-\eta R(n_i+n_j)\right]^{(\gamma)},  \\ 
\label{orbk}
{\hat K}_{ij}^{(\gamma)}&=&\frac{J}{2}\left[\eta R A_{ij}^{(\gamma)}
+\eta r B_{ij}^{(\gamma)}-\frac{1}{2}(1+\eta R)(n_i+n_j)\right]^{(\gamma)}.
\end{eqnarray}
Here $J=4t^2/U$, as usual. The coefficients $R=U/E_1=1/(1-3\eta)$ and 
$r=U/E_3=1/(1+2\eta)$ with $\eta=J_H/U$ take care of the $J_H-$multiplet 
structure. 

If we neglect the Hund's splitting of the excited states (consider 
$\eta\to 0$ limit), the Hamiltonian (\ref{model}) reduces to: 
\begin{equation}
H=J\sum_{\langle ij\rangle}
\frac{1}{2}({\vec S}_i\cdot {\vec S}_j+1)
\big({\vec\tau}_i\cdot {\vec\tau}_j+\frac{1}{4}n_i^{}n_j^{}\big)^{(\gamma)},
\label{pauli}
\end{equation}
where a constant energy of $-2J$ per site is neglected. 
This result should be compared with corresponding limit in the $d^1$ case, 
Eq.(\ref{Heta0}). One observes different spin structures:
$\frac{1}{2}({\vec S}_i\cdot {\vec S}_j+1)$ is obtained for vanadium
ions instead of $2({\vec S}_i\cdot {\vec S}_j+\frac{1}{4})$ for
spins one-half of Ti$^{3+}$. The difference in spin values 
can in fact be accounted for in general form  
as $({\vec S}_i\cdot {\vec S}_j+S^2)/2S^2$. It is important also to 
note that we have two electrons per V$^{3+}$ ion; one therefore has a 
different constraint equation for orbiton densities $n_{ia}+n_{ib}+n_{ic}=2$.

It is instructive to start again with a single bond along the $c$-axis. 
A crucial observation is that the lowest energy of $-J/2$ is 
obtained when the spins are {\it ferromagnetic}, and the 
orbitals $a$ and $b$ form a {\it singlet}, 
with $\langle {\vec\tau}_i\cdot {\vec\tau}_j\rangle^{(c)}=-\frac{3}{4}$. 
{\it Spin singlet$\otimes$orbital triplet} level is higher (at $-J/4$). 
This is in sharp contrast to the $S=1/2$ case, where the 
{\it spin singlet$\otimes$orbital triplet} and the 
{\it spin triplet$\otimes$orbital singlet} configurations 
are degenerate, resulting in a strong quantum resonance between them 
as it happens in titanates. Thus, ferromagnetic interactions are 
favored due to a local orbital singlet made
of $a$ and $b$ orbitals. Dominance of high spin configuration reflects
simply the fact that the spin part of the interaction, that is 
$({\vec S}_i\cdot {\vec S}_j+S^2)/2S^2$, is equal to 1 for a ferromagnetic
configuration, while vanishing in spin-singlet sector (as $-1/S$) in the 
limit of large spins. In order to form $ab-$orbital singlet on 
the bond along $c$ axis, the condition $n_{i}^{(c)}=n_{j}^{(c)}=1$ must 
be fulfilled (no $\vec\tau^{(c)}$ pseudospin can be formed otherwise). This
implies that the second electron on both sites has to go to an inactive
(that is $xy$) orbital. Thus we arrive at the following picture for the
superexchange bond in $c$ direction: ({\it i}) spins are aligned 
ferromagnetically, ({\it ii}) one electron on each site occupies 
either $a$ or $b$ orbital forming a {\it SU(2)} invariant orbital 
pseudospins that bind into the orbital singlet, ({\it iii}) the $xy$ orbital
obtains a stabilization energy of about $-J/2$ (the energy required 
to break $ab-$orbital singlet) and accommodates a remaining, second 
electron.

{\it Formation of one-dimensional orbital chains}.--- 
If the high spin state of the given pair is so stable, why does then a whole
crystal not form uniform ferromagnet? That is because each site has
two electrons, and an orbital that is inactive in particular 
(ferromagnetic bond) direction, 
induces in fact an antiferromagnetic coupling in the other two directions.
Thus spin interactions are strongly ferromagnetic (supported by orbital
singlets) in one direction, while the other bonds are antiferromagnetic.
As all directions are {\it a priori\/} equivalent in a cubic lattice, we 
again arrive at the problem of ``orbital frustration'' common to all
spin-orbital models on high-symmetry lattices. The solution of this problem 
here is as follows. As the spin-orbital resonance like in titanates is 
suppressed in the present case of large spin $S=1$, 
quantum energy can be gained mainly from the orbital 
sector. This implies that a particular classical spin configuration 
may be picked up which maximizes the energy gain from orbital fluctuations. 
Indeed, orbital singlets (with $n_{ia}+n_{ib}=1$) may form on the bonds 
parallel to the $c$-axis, exploiting fully the {\it SU(2)} orbital 
physics in one direction, while the second electron occupies 
the third $t_{2g}$ orbital ($n_{ic}=1$), controlling spin interactions in the 
$ab$-planes. Thus one arrives at spin order of the C-type
[ferromagnetic chains along $c$-axis which stagger within $ab$-planes],
which best comprises a mixture of ferromagnetic (driven by the orbital
singlets) and antiferromagnetic (induced by the electron residing 
on the static orbital) interactions. This is an analogy of the intermediate 
phase that we introduced for titanites above; here, it is much more
stable because of the large spin value. 

Once the C-type spin structure and simultaneous selection among the orbitals
(fluctuating $a,b$ orbitals, and stable $c$ orbital located at lower energy) 
is made by spontaneous breaking of the cubic symmetry, 
the superexchange Hamiltonian can be simplified.
We may set now $n_{ic}=1, n_{ia}+n_{ib}=1$, and introduce pseudospin
$\vec \tau$ which operates within the $(a,b)$ orbital doublet exclusively.

We focus first on orbital sector as quantum dynamics of the system
is dominated by the orbital pseudospins $\tau=\frac{1}{2}$
rather than by large spins $S=1$.
In the classical C-type spin state, $({\vec S}_i\cdot {\vec S}_j+1)$ is 
equal 2 along the $c$-direction while it is zero on 
$ab$-plane bonds. In this approximation, orbital interactions
in the model~(\ref{model}) are given by 
$(2{\hat J}_{ij}^{(c)}+{\hat K}_{ij}^{(c)})$ on $c$-axis bonds, 
while on $ab$-plane bonds only the ${\hat K}_{ij}^{(a,b)}$ term 
contributes which {\it is small}. Expressing $A_{ij}^{(\gamma)}$ and 
$B_{ij}^{(\gamma)}$ operators in Eqs.(\ref{orbj})--(\ref{orbk}) via 
pseudospins ${\vec\tau}$, one arrives at the following orbital 
Hamiltonian \cite{Kha01b}: 
\begin{equation}
H_{orb}=J_{orb}\sum_{\langle ij\rangle\parallel c}
(\vec \tau_i \cdot \vec \tau_j)
+J_{orb}^{\perp}\sum_{\langle ij\rangle\parallel (a,b)}\tau_i^z\tau_j^z~,
\label{Horbital}
\end{equation}
where $J_{orb}=JR$ and $J_{orb}^{\perp}=J\eta (R+r)/2$.
As their ratio is small, $J_{orb}^{\perp}/J_{orb}=\eta(1-5\eta r/2)$ 
(that is about only 0.1 for the realistic values of parameter
$\eta=J_H/U$ for vanadates), we obtain one-dimensional orbital 
pseudospin chains coupled only weakly to each other. 
Orbital excitations in the model (\ref{Horbital}) are mostly propagating
along $c$-chain directions. Their spectrum can be calculated, {\it e.g.}, 
within a linear spin-wave approximation, assuming a weak 
orbital order due to interchain coupling $J_{orb}^{\perp}$. One indeed 
obtains the one-dimensional {\it orbital-wave} spectrum \cite{Kha01b}: 
\begin{equation}
\omega_{\bf p}=\sqrt{\Delta^2+J^2_{orb}\sin^2 p_z},
\label{gapcaf}
\end{equation}
which shows the gap $\Delta=J\{\eta (R+r)[2R+\eta (R+r)]\}^{1/2}$ 
at $p_z=\pi$. The orbital gap $\Delta$ is small and grows with 
increasing Hund's coupling as $\propto J\sqrt{\eta}$.

Alternatively, one can use the Jordan-Wigner fermionic representation 
to describe quasi one-dimensional orbital dynamics, as suggested 
in Ref.\cite{Kha04a}. One obtains the 1D {\it orbiton} dispersion: 
\begin{equation}
\varepsilon_{\bf k}=\sqrt{h^2+J^2_{orb}\cos^2 k_z}, 
\end{equation}
where $h=4\tau J_{orb}^{\perp}$ is the ordering field stemming 
from interchain interactions. The staggered orbital order parameter 
$\tau=|\langle\tau^z_i\rangle|$, determined self-consistently from 
$\tau=\sum_k (h/2\varepsilon_k)\tanh(\varepsilon_k/2T)$, 
is small, and orbitals strongly fluctuate even at $T=0$. 

The underlying 1D-orbital dynamics have an important consequences
on spin interactions which control spinwave dispersions. 
In the spin sector, we obtain interactions $J_c({\vec S}_i\cdot {\vec S}_j)$ 
and $J_{ab}({\vec S}_i\cdot {\vec S}_j)$, with 
the spin-exchange constants following from Eq.(\ref{orbj}). 
The result is given by orbital pseudospin correlations: 
\begin{eqnarray}
\label{jcaf}
J_c&=&\frac{J}{2}\Bigl[(1+2\eta R)
\big\langle{\vec\tau}_i\cdot {\vec\tau}_j+\frac{1}{4}\big\rangle
-\eta r\big\langle{\vec\tau}_i\otimes{\vec\tau}_j+\frac{1}{4}\big\rangle
-\eta R\Bigl],   \\ \nonumber  
\label{jabaf}
J_{ab}&=&\frac{J}{4}\left[1-\eta (R + r)
 +(1+2\eta R-\eta r)\big\langle{\vec\tau}_i\otimes{\vec\tau}_j+
\frac{1}{4}\big\rangle\right].  
\end{eqnarray}
While in-plane antiferromagnetic couplings are mostly determined 
by the classical contribution of $xy$ orbitals (first term in $J_{ab}$), 
the exchange constant along the $c$-axis has substantial quantum 
contribution represented by the first term in $J_c$. This contribution 
is of negative sign due to the orbital singlet correlations along chains. 
The pseudospin expectation values in (\ref{jcaf}) can be estimated 
by using either the Jordan-Wigner fermionic-orbiton representation 
\cite{Kha04a} or within orbital-wave approximation \cite{Kha01b}. 
In both cases, one observes that the ferromagnetic coupling along 
the c-axis is strongly enhanced by orbital fluctuations. 
For a realistic values of $\eta \sim 0.12$, one obtains
$-J_c \sim J_{ab} \sim J/5$. This result is qualitatively different 
from that expected from the Goodenough-Kanamori rules. Indeed, in that 
classical picture with fully ordered orbitals one would obtain instead 
the smaller value $-J_c \sim 2\eta RJ_{ab} \sim J_{ab}/2$. 

Comparing all the coupling constants in both spin and orbital sectors, 
one observes that Heisenberg-like orbital dynamics has the largest 
energy scale, $J_{orb}=JR$, thus dominating the physics of the 
present spin-orbital model. The overall picture is that 
``cubic frustration'' is resolved here by the formation 
of the orbital chains with Heisenberg dynamics, from which 
a large quantum energy is gained. This is similar to the  
order-from-disorder scenario for the $e_g$ spin-orbital model 
(\ref{HAM2}), where classical $3z^2-r^2$ orbital order results
in quasi one-dimensional spin chains. In the present $t_{2g}$ orbital
model with large classical spins, the role of spins and orbitals are just 
interchanged. 

As argued in Ref.\cite{Kha01b}, the above scenario may explain 
the $C$-AF type spin order in LaVO$_3$~\cite{Miy02}. A structural transition 
that follows the onset of magnetic order in this compound is also 
natural in the superexchange model. Indeed, C-type spin ordering and 
formation of the pseudospin orbital chains are intimately connected and 
support each other. Thus, the ordering process in the orbital sector --- 
the stabilization of $xy$ orbital which absorbs one electron, 
and lifting the remaining degeneracy of $xz/yz$-doublet via 
quantum fluctuations --- occurs cooperatively with the evolution 
of the $C$-type magnetic order. In more formal terms, the classical order 
parameters $Q_{orb}=(2n_c-n_a-n_b)$ ($xy$-orbital selection) 
and $Q_{sp}=(\langle{\vec S}_i\cdot {\vec S}_j\rangle_{c}
-\langle{\vec S}_i\cdot {\vec S}_j\rangle_{ab})/2$ (spin-bond selection) act 
cooperatively to break the cubic symmetry, while the energetics is driven 
by a quantum energy released by $xz/yz$ orbital fluctuations. 
Obviously, this picture would break down {\it if} the JT-coupling 
dominates --- nothing but a conventional 3D-classical ordering 
of $xz/yz$-doublets at some $T_{str}$, independent of spin correlations, 
would take place in that case. 

A pronounced anisotropy and temperature dependence of the optical 
conductivity observed in LaVO$_3$~\cite{Miy02} also find a {\it quantitative} 
description~\cite{Kha04a} within this theory, in which quantum orbital 
fluctuations play the key role. Interestingly, the JT picture has also 
been applied to the problem, see Ref.\cite{Mot03}. Based on first-principle 
calculations, a JT binding energy $E_{JT}\sim 27$~meV has been obtained. 
Consequently, a large orbital splitting ($=4E_{JT}$) suppresses 
the quantum nature of orbitals and, as a result, one obtains that the 
optical spectral weghts along the $c$ and $a,b$ directions are 
almost the same, $I_c/I_{ab}\simeq 1$, contradicting the experiment 
which shows strong polarization dependence. 
In our view, the optical experiments~\cite{Miy02} clearly indicate 
that JT coupling is much smaller than estimated in Ref.\cite{Mot03}, 
and are thus not sufficient to lock-in the orbitals in LaVO$_3$. At this
point, we are faced again with a problem of the principal importance: 
Why do the first-principle calculations always predict a large 
orbital splittings? should those numbers be directly used as an input 
parameters in the many-body Hamiltonians, suppressing thereby 
all the interesting orbital physics? These questions remain puzzling. 

Based on the above theory, and by analogy with titanites, we expect that 
the Raman light scattering on orbital fluctuations in vanadates should 
be visible as a {\it broad} band in the energy range of two-orbiton 
excitations, $2J_{orb}$, of the quantum orbital chains. 
Given a single-orbiton energy scale $J_{orb}=JR\sim 60-70$~meV 
(which follows from the fit of optical \cite{Miy02} 
and neutron scattering \cite{Ulr03} data), 
we predict a broad Raman band centered near $\sim 120$~meV. This energy 
is smaller than in titanites because of the low-dimensionality of orbital 
dynamics, and falls in the range of two-phonon spectra. However, 
the orbital-Raman band is expected to have very large width 
(a large fraction of $J_{orb}$), as orbitons strongly scatter 
on spin fluctuations (as well on phonons, of course). More specifically, we 
observe that both thermal and quantum fluctuations of the spin-bond operator, 
$({\vec S}_i\cdot{\vec S}_j+1)$ in Eq.(\ref{model}) or (\ref{pauli}), 
introduce an effective disorder in the orbital sector. In other words, 
the orbital Hamiltonian (\ref{Horbital}) obtains a strong  
bond-disorder in its coupling constants, hence the expected 
incoherence and broadness of the orbital Raman response. 
This feature, as well as a specific temperature and polarization 
dependences should be helpful to discriminate two-orbiton band 
from two-phonon (typically, sharply structured) response. 
On the theory side, the light scattering problem in the full spin-orbital 
model (\ref{model}) need to be solved in order to figure out 
the lineshape, temperature dependence, {\it etc}. Alternatively, 
a crystal-field and/or first-principle calculations would be helpful, 
in order to test further the "lattice-distortion" picture 
for vanadates, --- in that case, the Raman-band frequencies would be 
dictated by the on-site level structure. 

Let us conclude this section, devoted to the spin-orbital models: 
Orbital frustration is a generic feature of the superexchange 
interactions in oxides, and leads to a large manifold 
of competing states. Reduction of the effective dimensionality by 
the formation of quantum spin- or orbital-chains (depending on which 
sector is "more quantum") is one way of resolving the frustrations. 
In $e_g$ spin-orbital models, the order-from-disorder mechanism completes 
the job, generating a finite orbital gap below which a classical 
description becomes valid. 
In superexchange models for titanites, where both spins and orbitals 
are of quantum nature, a composite spin-orbital bound states may develop 
in the ground state, lifting the orbital degeneracy {\it without breaking} 
the cubic symmetry. The nature of such a quantum orbital-liquid 
and its elementary excitations is of great theoretical interest, 
regardless to which extend it is "spoiled" in real materials. 
The interplay between the SE interactions, dynamical JT coupling, 
and extrinsic lattice distortions is a further step towards a quantitative 
description of the electronic phase selection in titanites.     

\section{Competition between superexchange and lattice effects: YVO$_3$} 
As an example of the phase control by a competing superexchange, lattice, 
temperature and doping effects, we consider now a slightly modified version 
of the $S=1$ model (\ref{model}) for vanadates, by adding there the following 
Ising-like term which operates on the bonds along $c$ direction\cite{Kha01b}: 
\begin{equation}
H_V=-V\sum_{\langle ij\rangle\parallel c}\tau^z_i\tau^z_j. 
\label{ising}
\end{equation}
This describes a ferromagnetic ($V>0$) orbital coupling, which competes 
with the orbital-Heisenberg $J_{orb}-$term in the Hamiltonian 
(\ref{Horbital}). Physically, this term stands for the effect of GdFeO${_3}$ 
type distortions including A--O covalency, which prefer a stacking 
("ferromagnetic" alignment) of orbitals along $c$ direction\cite{Miz99}. 
This effect gradually increases as the size of A-site 
ion decreases and may therefore 
be particularly important for YVO$_3$. Interesting to note that 
the $V-$term makes the orbital interactions of $xyz$-type with $z$-coupling 
being smaller, hence driving the orbital sector towards a more disordered 
$xy$ model limit (motivating the use of Jordan-Wigner representation 
for orbitals~\cite{Kha04a}). However, when increased above a critical 
strength, the $V-$term favors a classically ordered orbitals. 

Three competing phases can be identified for the modified, 
(\ref{model}) plus (\ref{ising}), model. ({\it i}) The first one, 
which is stable at moderate $V<J$, was that just considered above: 
$C$-type spin order with fluctuating orbitals (on top of a very weak 
staggered order) as in LaVO$_3$. This state is driven by a cooperative 
action of orbital-singlet correlations and $J_H$--terms. 
({\it ii}) At large $V>J$, the orbitals order ferromagnetically 
along the $c$-axis, enforcing the $G$-type spin order, 
as observed in YVO$_3$. ({\it iii}) Yet there is a third state, 
which is "hidden" in the SE-model and may 
become a ground state when both $J_H$ and $V$ are below some critical values. 
This is a valence bond solid (VBS) state, which is derived by a dimerization 
of the Heisenberg orbital chains, by using the spin-bond operators 
$({\vec S}_i\cdot{\vec S}_j+1)$ as an "external" dimerization field. 
(We already mentioned this operator as a "disorder field" for 
the orbital excitations). 

Specifically, consider the limit of $\eta=0, V=0$, where we are left 
with the model (\ref{pauli}) alone. It is easy to see that we can 
gain more orbital quantum energy by choosing the following 
spin structure: On every second $c$-axis bond, spins are ferromagnetic,  
$({\vec S}_i\cdot {\vec S}_j+1)=2$, while they are antiparallel in all other
bonds giving $({\vec S}_i\cdot {\vec S}_j+1)\sim 0$. Consequently, 
the orbital singlets are fully developed on ferro-spin bonds, gaining 
quantum energy $-J/2$ and supporting the high-spin state assumed. 
On the other hand, the orbitals are completely decoupled on AF-spin bonds.   
As a result, the expectation value $\langle\vec\tau_i \cdot \vec\tau_j\rangle$
vanishes on these ``weak'' bonds, thus the spin exchange constant
$J_s^{weak}=J/8$ (consider Eq.~(\ref{jcaf}) for uncorrelated
orbitals and the $\eta=0$ limit) turns out to be positive, 
consistent with antiferromagnetic spin alignment on these bonds.
Such a self-organized spin-orbital dimer state is lower 
in energy [$-J/4$ per site] than a uniform Heisenberg orbital chains 
with $C$-type spin structure [$(1/2-\ln2)J \simeq -0.19J$]. Thus, 
the VBS state (building blocks are the decoupled orbital dimers 
with total spin 2) is a competing state at small $\eta$ values, as 
has been noticed by several authors \cite{She02,Ulr03,Sir03,Hor03,Miy04}. 

\subsection{Phase diagram}
Interplay between the above three spin-orbital structures 
has been investigated numerically in Ref.\cite{Miy04} within the DMRG method. 
The thermodynamic properties and interesting evolution of correlation 
functions have also been studied by using the finite-temperature 
version of DMRG \cite{Sir03,Ulr03}. Though these methods are restricted 
to the one-dimensional version of the model, they give rigorous 
results and are in fact quite well justified, because the essential 
physics is governed by strong dynamics within the $\sim$1D 
orbital chains, while a weak interchain couplings can be 
implemented on a classical mean-field level \cite{Sir03}. 
\begin{figure}[htb]
\begin{center}
\epsfxsize=80mm
\centerline{\epsffile{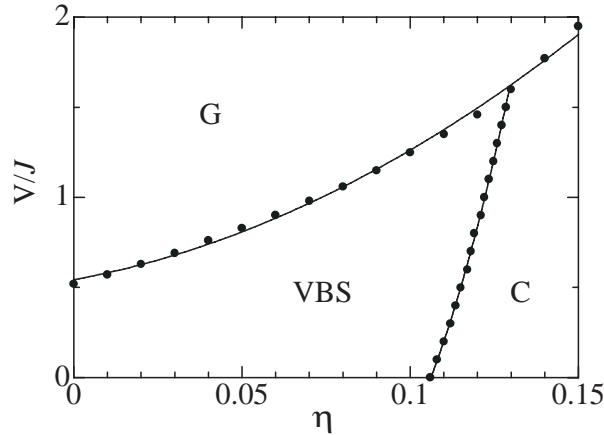}}
\end{center}
\caption{
Phase diagram in the $\eta$-$V$ plane. 
The VBS state consists of orbital-singlet dimers with spin 2, and the spins 
of different dimers are weakly coupled antiferromagnetically. 
In the phase $C$, a uniform orbital-chain is restored, 
while the spins are aligned ferromagnetically along the chain. 
The phase $G$ is the spin-AF/orbital-ferromagnetic state, stabilized 
by the Ising interaction between the orbitals originating from the 
GdFeO${_3}$-type distortions. All the phase transitions are of first-order. 
(After Ref.~\protect\cite{Miy04}). 
}
\label{fig3}
\end{figure}

The ground state phase diagram in the $\eta$-$V$ plane, obtained from the 
DMRG study\cite{Miy04} is shown in Fig.\ref{fig3}. There are three 
distinct phases in this figure. For small $\eta$ and $V$, the VBS state 
is stabilized, which is driven either to the 
orbital-ferromagnetic/spin-AF phase (called $G$) with the increase of $V$, 
or to the orbital-AF/spin-ferromagnetic one (called $C$) with the increase 
of the Hund's coupling $\eta$. The critical value $\eta_c(V=0)\simeq 0.11$ 
for the latter transition perfectly agrees with the earlier result inferred 
from the finite-temperature DMRG-study \cite{Sir03}, and is just slightly 
below the realistic values in vanadates \cite{Miz96}. This indicates 
the proximity of VBS state in vanadates, on which we elaborate later-on.   
Note that when the Hund's coupling is slightly above the critical 
value, the ground state is spin-ferromagnetic for small $V$, 
but for intermediate values of $V$ the VBS-state is stabilized. 
Stabilization of the {\it orbital disordered} state by 
finite $V$ interaction (induced by {\it lattice distortion}) 
is a remarkable result. The physics behind this observation is that ---  
as already mentioned --- the interaction (\ref{ising}) introduces 
a frustration into the orbital sector, competing with antiferro-type 
alignment of orbitals in the SE model. 

The $V-$driven phase transition from the spin-$C$ to spin-$G$ ground state, 
obtained in these calculations describes physically 
why an increased GdFeO$_3$-type distortion promotes 
a spin-staggering along the $c$ direction in YVO$_3$, while less distorted 
LaVO$_3$ has the spin-$C$ structure. This study also suggests that 
$V\sim J$ in vanadates. As $J\sim 40$~meV in these compounds 
\cite{Ulr03,Kha04a}, we suspect that the effect of GdFeO$_3$-type 
distortions on $t_{2g}$ orbitals is roughly of this scale in perovskites 
in general, including titanites. Being comparable with the SE scale 
$J_{orb}\sim 60$~meV in vanadates, the $V-$ term is important to stabilize 
the spin-$G$ phase, but it is not sufficient to lock-in the orbitals 
in titanates, where {\it the three dimensional} exchange fluctuations 
are characterized by larger energies of the order of 
$W_{orb}\sim 120$~meV (see previous Section). 

\subsection{Entropy driven spin-orbital dimerization} 
YVO$_3$ is a remarkable material, in the sense that its spin-$G$ type 
ground state is very fragile and readily changes to the spin-$C$ 
state via the phase transition, driven either by temperature (at 77~K) or 
by small doping \cite{Fuj05}. This finds a natural explanation within the
present theory in terms of competing phases, assuming that YVO$_3$ is 
located in the phase $G$ near the tricritical point (see the phase diagram
in Fig.\ref{fig3}), and thus close to the spin-$C$ and VBS states. 

This view is strongly supported by the neutron scattering data of 
Ref.\cite{Ulr03}. This experiment revealed several anomalies, indicating 
that the spin-$C$ phase above 77~K is itself very unusual: 
({\it i}) substantial modulations of the spin couplings along 
the $c$ direction, ({\it ii}) ferromagnetic interactions are stronger than  
$ab-$plane AF ones, ({\it iii}) anomalously small ordered moment, which 
({\it iv}) selects the $ab-$plane (different from the easy $c$ axis  
found just below 77~K or in the $C$-phase of LaVO$_3$). 
All these features have coherently been explained in Ref.\cite{Ulr03} in 
terms of underlying quantum dynamics of the spin-orbital chains and their 
{\it entropy-driven} dimerization \cite{Sir03}. Physics behind the last 
point is as follows. 

Because the Hund's coupling parameter is close to the critical one, 
there is strong competition between uniform ($C$) and dimerized (VBS) 
states, and this may affect thermodynamic properties for entropy reasons. 
The point is that the dimer state contains ``weak'' spin bonds: 
spin interaction between different dimers is small when $\eta$ is 
near the critical value (when the dimerization amplitude is not 
saturated, weak bonds are ferromagnetic but much weaker than strong 
ferromagnetic bonds within the dimers). Therefore, the spin entropy of 
an individual dimer with total spin 2, that is {\it ln}~5, is released. 
The gain of spin entropy due to the dimerization lowers the free 
energy $F=\langle H \rangle-TS$ and may stabilize a dimerized 
state with alternating weak and strong ferro-bonds 
along $c$-axis. In other words, the dimerization of spin-orbital chains
occurs due to the {\it orbital Peierls effect}, in which thermal
fluctuations of the spin bond-operator $({\vec S}_i\cdot {\vec S}_j)_{c}$
play the role of lattice degrees of freedom, 
while the critical behavior of the Heisenberg-like orbital chains 
is a driving force. As the dimerization is of electronic origin, 
and is not as complete as in the VBS state itself, concomitant 
lattice distortions are expected to be small. 

These qualitative arguments have been substantiated by numerical 
studies using the finite-temperature DMRG method~\cite{Sir03,Ulr03}. 
The explicit calculations of the entropy, evolution of the dimer correlations 
upon increasing temperature, and anomalous behavior of spin and orbital
susceptibilities can be found in these papers. 

\subsection{Doping control}
The energy difference between $G$ and $C$ type spin orderings 
in YVO$_3$ is very small, $E_C-E_G \sim 0.1 J \sim 4$~meV only  
\cite{Kha01b,Miy04} for realistic values of $\eta$ and $V$. 
Therefore, the $G$-type ground state of YVO$_3$ can easily change 
to the $C$-type upon small perturbations. For instance, pressure 
may reduce the GdFeO$_3$ distortions hence the $V-$interaction, triggering 
the first-order phase transition described above. Injection of the 
charge carries is the another possibility, which we discuss now 
(see also Ref.\cite{Ish05}).     

We need to compare a kinetic energy gains in above-mentioned competing phases. 
It is easy to see that the charge-carriers strongly favor spin-$C$ phase, 
as they can freely move along the ferromagnetic spin chains in the 
$c$ direction of the $C$ phase, while a hole-motion is frustrated 
in the spin-$G$ phase in all directions. In more detail, the $ab$-plane 
motion is degraded in both phases equally by a classical AF order via the 
famous "double-exchange factor" $\cos\frac{\theta}{2}$ with $\theta=\pi$ 
for antiparallel spins. Thus, let us focus on the $c$ direction. 
In the $G$-phase, a hole-motion is disfavored again by antiparallel 
spins ($\theta_c=\pi$), but the spin-$C$ phase with $\theta_c=0$ is 
not "resistive" at all. Now, let us look at the orbital sector. The orbitals 
are fully aligned in the spin-$G$ phase and thus introduce no frustration. 
Our crucial observation is that the orbitals are not resistive to 
the hole motion in the spin-$C$ phase, too. 
The point is that a doped-hole in the spin-$C$ phase can be regarded as 
a {\it fermionic-holon} of the quasi-1D orbital chains, and its motion 
is not frustrated by orbitals at all. This is because of the 
orbital-charge separation just like in case of a hole motion in 
1D Heisenberg chains\cite{Oga90}. As a result, the $c$ direction is fully  
transparent in the spin-$C$ phase (both in spin and orbital sectors), 
and doped holes gain a kinetic energy $K=2tx$ per site at small $x$. 
(On contrast, a hole motion is strongly disfavored in spin-$G$ phase 
because of AF-alignment of large spins $S=1$, as mentioned). 
Even for the doping level as small as $x=0.02$, this gives an 
energy gain of about 8~meV (at $t\sim 0.2$~eV) for the $C$ phase, 
thus making it {\it lower} in energy than the $G$-phase. 
This is presicely what is observed \cite{Fuj05}. The underlying 
quantum orbital chains in the spin-$C$ phase \cite{Kha01b} are of crucial 
importance here: a {\it static} configuration of the staggered 
$xz/yz$ orbitals would discourage a hole-motion. 
In other words, fluctuating orbitals not only support the ferromagnetic 
spins, but also well accommodate doped holes because 
of orbital-charge separation. The {\it quantum orbitals} and a doping induced 
{\it double-exchange} act cooperatively to pick up the spin-$C$ phase as 
a ground state. 

The "holon-like" behavior of the doped-holes implies a quasi one-dimensional 
charge transport in doped vanadates, namely, a much larger and strongly 
temperature dependent low-energy optical intensity (within the Mott-gap) 
along the $c$ axis, in comparison to that in the $ab$-plane polarization. 
The situation would be quite different in case of classical JT orbitals: 
A staggered 3D-pattern of static orbitals frustrates 
hole motion in all three directions, and only a moderate anisotropy 
of the low-energy spectral weights is expected. This way, optical 
experiments in a doped compounds have a potential to discriminate 
between the classical JT picture and quantum orbitals, just like 
in the case of pure LaVO$_3$ as discussed above. 

The doping-induced transition from the $G$ to the $C$ phase must be of the 
first order, as the order parameters and excitations of the $G$ phase 
in both spin and orbital 
sectors hardly change at such small doping levels. Specifically, AF spin 
coupling $J_c$ in the $G$ phase may obtain a double-exchange ferromagnetic 
correction $J_{DE}< K/4S^2=tx/2$ due to the local vibrations of 
doped holes, which is about 2~meV only at the critical doping $x\sim0.02$.  
This correction is 
much smaller than $J_c\simeq6$~meV \cite{Ulr03} of undoped YVO$_3$. 
Regarding the orbitals, they are perfectly aligned "ferromagnetically" 
along the $c$ axis in the ground state of YVO$_3$, satisfying the GdFeO$_3$ 
distortion; we find that this is not affected by 1-2\% doping at all. 
Our predictions are then as follows: The staggered spin moment and spinwave
dispersions of the $G$ phase remain barely unrenormalized upon doping, 
until the system suddenly changes its ground state. On the other hand, 
we expect that doping should lead to sizable changes in the properties 
of spin-$C$ phase, because of the positive interplay between the 
"holons" formed and underlying orbital quantum physics. Our predictions are 
opposite to those of Ref.\cite{Ish05}; this gives yet another 
opportunity to check experimentally (by means of neutron scattering) 
which approach --- SE quantum picture or classical JT orbitals --- 
is more appropriate for the spin-$C$ phase of vanadates. 

A more intriguing question is, however, whether the doping induced 
$C$-phase of YVO$_3$ is also dimerized as in the undoped YVO$_3$ itself, 
or not. Theoretically, this is well possible at small doping, 
and, in fact, we expect that the orbital Peierls effect should cooperate 
with a conventional fermionic one, stemming from the hopping-modulation 
of doped carriers along the underlying orbital chains. 

Summarizing this section, we conclude that the $t_{2g}$ spin-orbital model 
with high spin values shows an intrinsic tendency towards dimerization. 
Considered together with lattice induced interactions, 
this leads to a several low-energy competing many-body states, 
and to the "fine" phase-selection by temperature, distortions 
and doping. One of these phases is the spin-$G$ phase with classical
orbitals; it is well isolated in the Hilbert space and hence may 
show up only via a discontinuous transition. 

\section{Lifting the orbital degeneracy by spin-orbit coupling: Cobaltates}
It is well known \cite{Kan56,Goo63,Kug82} that in $t_{2g}$ orbital 
systems a conventional spin-orbit coupling 
$\lambda(\vec S \cdot \vec l)$ may sometimes play a crucial 
role in lifting the orbital degeneracy, particularly in case of 
late-transition metal ions with large $\lambda$. When it dominates 
over the superexchange and weak orbital-lattice interactions, 
the spin and orbital degrees of freedom are no longer separated, 
and it is more convenient to formulate the problem in terms 
of the total angular momentum. Quite often, the spin-orbit ground state 
has just the twofold Kramers degeneracy, and low-energy magnetic 
interactions can be mapped on the pseudospin one-half Hamiltonian. 
This greatly reduces the (initially) large spin-orbital Hilbert space. 
But, as there is "no free lunch", the pseudospin Hamiltonians 
obtain a nontrivial structure, because the bond directionality 
and frustrations of the orbital interactions are transfered 
to the pseudospin sector via the spin-orbital unification. 
Because of its composite nature, ordering of pseudospin necessarily 
implies both $\vec S$- and $\vec l$-orderings, and the "ordered" orbital 
in this case is a complex function. Via its orbital component, 
the pseudospin order pattern is rather sensitive to the lattice geometry. 

Experimental indications for such physics are as follows: 
({\it i}) a separate JT-structural transition is suppressed but a  
large magnetostriction effects occur upon magnetic ordering;   
({\it ii}) effective $g-$values in the ground state may deviate 
from the spin-only value and are anisotropic in general;  
({\it iii}) magnetic order may have a complicated nontrivial structure 
because of the non-Heisenberg symmetry of pseudospin interactions. 
As an example, the lowest level of Co$^{2+}$ ions 
($t^5_{2g}e^2_g$, $S=3/2$, $l=1$) in a canonical Mott insulators 
CoO and KCoF$_3$ is well described by a pseudospin one-half~\cite{Kan56}. 
Low-energy spin waves in this pseudospin sector, which are separated 
from less dispersive local transitions to the higher levels with different  
total momentum, have been observed in neutron scattering 
experiments in KCoF$_3$~\cite{Hol71}.     

In this section, we apply the pseudospin approach to study the 
magnetic correlations for the triangular lattice of CoO$_2$ planes, 
in which the Co$^{4+}(t^5_{2g})$ ions interact via the $\sim 90^{\circ}$ 
Co-O-Co bonds. We derive and analyze the exchange interactions 
in this system, and illustrate how a spin-orbit coupling leads to unusual 
magnetic correlations. This study is motivated by recent interest in  
layered cobalt oxides Na$_x$CoO$_2$ which have a complex phase 
diagram including superconductivity, charge and magnetic 
orderings~\cite{Foo03}.   

The CoO$_2$ layer consists of edge sharing CoO$_6$ octahedra 
slightly compressed along the trigonal ($c \parallel$ [111]) axis.  
Co ions form a 2D triangular lattice, sandwiched by oxygen layers. 
It is currently under debate~\cite{Lee05}, whether the 
undoped CoO$_2$ layer (not yet studied experimentally) 
could be regarded as Mott insulator or not. Considering the 
strongly correlated nature of the electronic states in doped cobaltates 
as is supported by numerous magnetic, transport and photoemission 
measurements, we assume that the undoped CoO$_2$ plane is on the insulating 
side or near the borderline. Thus the notion of spin-charge energy scale 
separation and hence the superexchange picture is valid at least locally. 
  
\subsection{Superexchange interaction for pseudospins}
A minimal model for cobaltates should include the orbital degeneracy 
of the Co$^{4+}-$ion \cite{Kos03}, where a hole in the 
$d^5(t_{2g})$ shell has the freedom to occupy one out of three  
orbitals $a=d_{yz}$, $b=d_{xz}$, $c=d_{xy}$. The degeneracy is partially 
lifted by trigonal distortion, which stabilizes $A_{1g}$ electronic 
state $(a+b+c)/\sqrt{3}$ over $E'_g$-doublet 
$(e^{\pm i\phi}a+e^{\mp i\phi}b+c)/\sqrt{3}$ (hereafter $\phi=2\pi/3$): 
\begin{equation}
H_{\Delta}=\Delta [n(E'_g)-2n(A_{1g})]/3. 
\end{equation}
The value of $\Delta$ is not known; Ref.\cite{Kos03} estimates it 
$\Delta\sim$25~meV. Physically, $\Delta$ should be sample dependent 
thus being one of the control parameters. We will later see that 
magnetic correlations are indeed very sensitive to $\Delta$.   
\begin{figure} 
\epsfxsize=0.80\hsize \epsfclipon \centerline{\epsffile{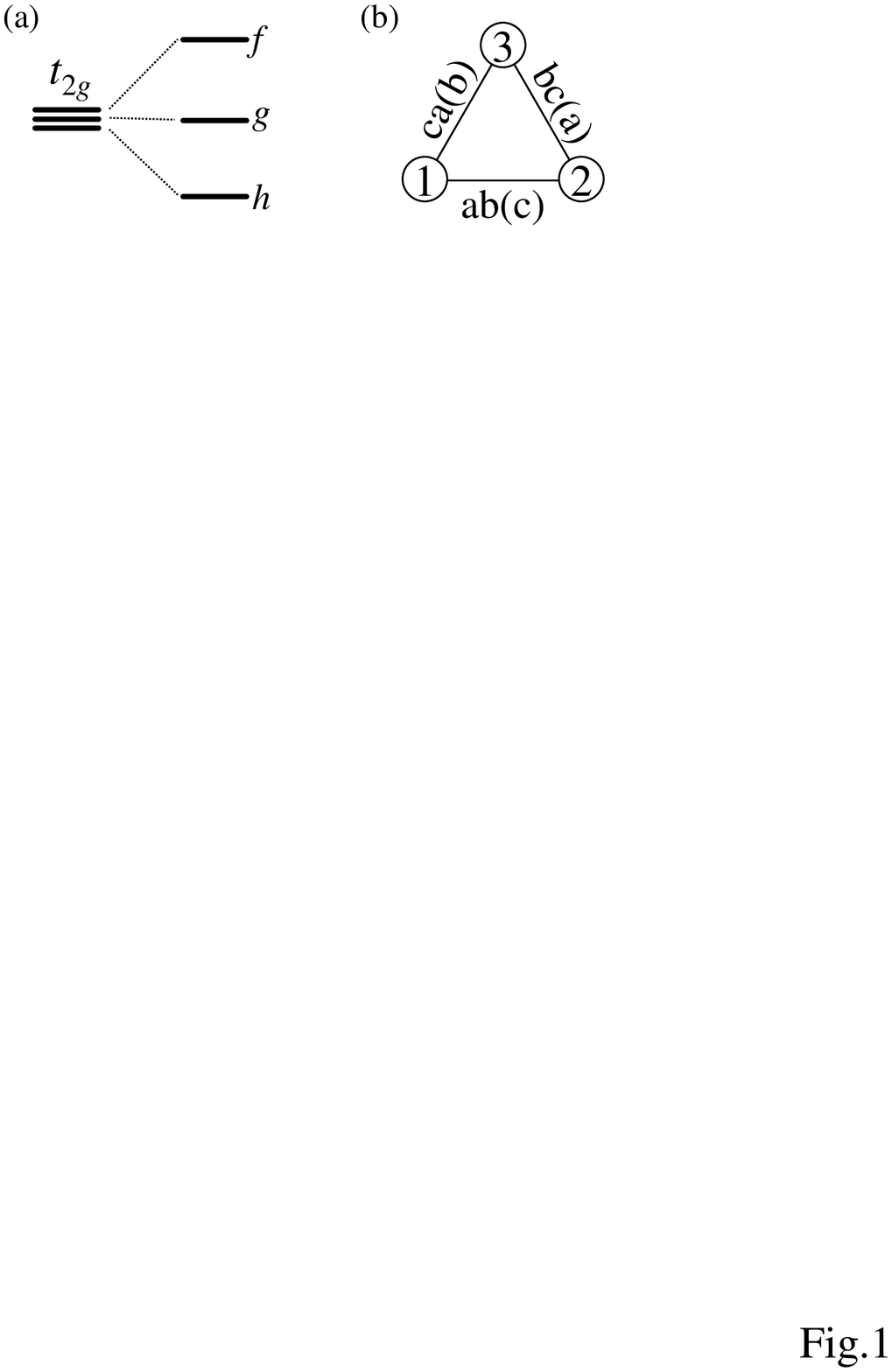}}
\caption{
${\bf (a)}$ The $t_{2g}$-orbital degeneracy of Co$^{4+}(d^5)$-ion is 
lifted by trigonal distortion and spin-orbit interaction.  
A hole with pseudospin one-half resides on the Kramers $f$-doublet.  
Its wavefunction contains both $E'_g$ and $A_{1g}$ states 
mixed up by spin-orbit coupling. 
${\bf (b)}$ Hopping geometry on the triangular lattice of Co-ions. 
$\alpha \beta (\gamma)$ on bonds should be read as $t_{\alpha \beta}=t$, 
$t_{\gamma \gamma}=-t'$, and $\alpha,\beta,\gamma \in \{a,b,c\}$ 
with $a=d_{yz}$, $b=d_{xz}$, $c=d_{xy}$. The orbital nondiagonal  
$t$-hopping stems from the charge-transfer process via oxygen ions,  
while $t'$ stands for the hopping between the same orbitals 
due to either their direct overlap or via two intermediate oxygen ions 
involving $t_{pp}$. (After Ref.~\protect\cite{Kha04b}).    
}
\label{fig4}
\end{figure}

In terms of the effective angular momentum $l=1$ of the $t_{2g}$-shell, 
the functions $A_{1g}$ and $E'_g$ correspond to 
the $|l_z=0\rangle$ and $|l_z=\pm 1\rangle$ states, respectively. 
Therefore, a hole residing on the $E'_g$ orbital doublet will experience 
an unquenched spin-orbit interaction 
$H_{\lambda}=-\lambda({\vec S}\cdot {\vec l})$. 
The coupling constant $\lambda$ for a free Co$^{4+}$ ion is 
650 cm$^{-1}\approx 80$meV \cite{Fig00} (this may change in a solid 
due to the covalency effects). 
  
The Hamiltonian $H=H_{\Delta}+H_{\lambda}$ is diagonalized by the following 
transformation\cite{Kha04b}:
\begin{eqnarray}
\label{eq1}
\alpha_{\sigma}=i[c_{\theta}e^{-i\sigma \psi_{\alpha}}f_{-\bar\sigma}+ 
is_{\theta}f_{\bar\sigma}+e^{i\sigma \psi_{\alpha}}g_{\bar\sigma}+ 
s_{\theta}e^{-i\sigma \psi_{\alpha}}h_{-\bar\sigma}-
ic_{\theta}h_{\bar\sigma}]/\sqrt{3}~, 
\end{eqnarray}
where $c_{\theta}=\cos\theta, s_{\theta}=\sin\theta$, 
$\alpha=(a,b,c)$ and $\psi_{\alpha}=(\phi, -\phi, 0)$,
correspondingly. The angle $\theta$ is determined from  
$\tan{2\theta}=2\sqrt{2}\lambda/(\lambda + 2\Delta)$. As a result, one 
obtains three levels, 
$f_{\bar\sigma}, g_{\bar\sigma}, h_{\bar\sigma}$ [see Fig.\ref{fig4}(a)],  
each of them are Kramers doublets with pseudospin one-half $\bar\sigma$. 
The highest, $f$-level, which accommodates a hole in $d^5$ configuration, 
is separated from the $g$-level by   
$\varepsilon_f-\varepsilon_g=
\lambda+\frac{1}{2}(\lambda/2+\Delta)(1/\cos{2\theta}-1)$. 
This splitting is $\sim 3\lambda/2$ at $\lambda \gg \Delta$, and 
$\sim\lambda$ in the opposite limit. 
It is more convenient to use a hole representation, in which 
the level structure is reversed such that $f$-level is the lowest one. 
It is important to observe that the pseudospin $f_{\bar\sigma}$ states 
\begin{eqnarray}
\label{eq2}
|\bar\uparrow \rangle_f&=&ic_{\theta}|+1,\downarrow\rangle - 
s_{\theta}|0,\uparrow\rangle, \nonumber \\
|\bar\downarrow \rangle_f&=&ic_{\theta}|-1,\uparrow\rangle - 
s_{\theta}|0,\downarrow\rangle 
\end{eqnarray}
are coherent mixture of different orbital and spin states. This will 
have important consequences for the symmetry of intersite interactions.       

We assume the hopping Hamiltonian suggested by the edge-shared structure: 
\begin{equation}
H_t^{ij}=t(\alpha^{\dagger}_{i\sigma}\beta_{j\sigma}+
\beta^{\dagger}_{i\sigma}\alpha_{j\sigma})-
t'\gamma^{\dagger}_{i\sigma}\gamma_{j\sigma} + h.c.~, 
\end{equation}
where $t=t_{\pi}^2/\Delta_{pd}$ originates from $d$-$p$-$d$ process via the  
charge-transfer gap $\Delta_{pd}$, and $t'>0$ is given either by direct 
$d$-$d$ overlap or generated by indirect processes like $d$-$p$-$p$-$d$.  
On each bond, there are two orbitals active in the $d$-$p$-$d$ process, 
while the third one is transfered via the $t'$-channel [Fig.\ref{fig1}(b)]. 
For simplicity, we assume $t'<t$ and neglect $t'$ for a while.   

There are three important superexchange processes:   
({\bf a}) Two holes meet each other at the Co-site; the excitation 
energy is $U_d$. 
({\bf b}) Two holes meet each other at an intermediate oxygen-site; 
the excitation energy $2\Delta_{pd}+U_p$. 
({\bf c}) The oxygen electron is transfered to an unoccupied $e_g$ shell   
and polarizes the spin of the $t_{2g}$ level via the Hund's interaction, 
$-2J_H({\vec s}_e \cdot {\vec s}_t)$. This process   
is important because the $e_g-p$ hopping integral $t_{\sigma}$  
is larger than $t_{\pi}$. The process (b), 
termed "a correlation effect" by Goodenough~\cite{Goo63}, 
is expected to be stronger than contribution (a), 
as cobaltates belong to the charge-transfer insulators~\cite{Zaa85}.    

These three virtual charge fluctuations give the following contributions:  
\begin{eqnarray} 
\label{A}
({\bf a}):\;\;\;\; 
&\;&A({\vec S}_i\cdot{\vec S}_j+1/4)(n_{ia}n_{jb}+n_{ib}n_{ja}+
a^{\dagger}_ib_ia^{\dagger}_jb_j+b^{\dagger}_ia_ib^{\dagger}_ja_j), \\
({\bf b}):\;\;\;\; 
&\;&B({\vec S}_i\cdot{\vec S}_j-1/4)(n_{ia}n_{jb}+n_{ib}n_{ja}), \\
({\bf c}):\;\;\;\; 
&-&C({\vec S}_i\cdot{\vec S}_j)(n_{ic}+n_{jc}), 
\end{eqnarray} 
where $A=4t^2/U_d$, $B=4t^2/(\Delta_{pd}+U_p/2)$ and $C=BR$, with 
$R\simeq(2J_H/\Delta_{pd})(t_{\sigma}/t_{\pi})^2$; 
$R\sim 1.5-2$ might be a realistic estimation. [As usual, the orbitals 
involved depend on the bond direction, and the above equations refer 
to the 1-2 bond in Fig.\ref{fig4}(b)]. The first, $A-$contribution 
can be presented in a {\it SU(4)} form (\ref{Heta0}) like in 
titanites and may have ferromagnetic as well as antiferromagnetic 
character in spin sector depending on actual orbital correlations. 
While the second (third) contribution is definitely
antiferromagnetic (ferromagnetic). As the constants 
$A\sim B\sim C\sim 20-30$~meV are smaller than spin-orbit splitting, 
we may now {\it project} the above superexchange Hamiltonian 
{\it onto the lowest Kramers level} $f$, obtaining the effective 
low-energy interactions between the pseudospins one-half $\vec S_f$. 
Two limiting cases are presented and discussed below. 

\begin{figure} 
\centerline{\epsffile{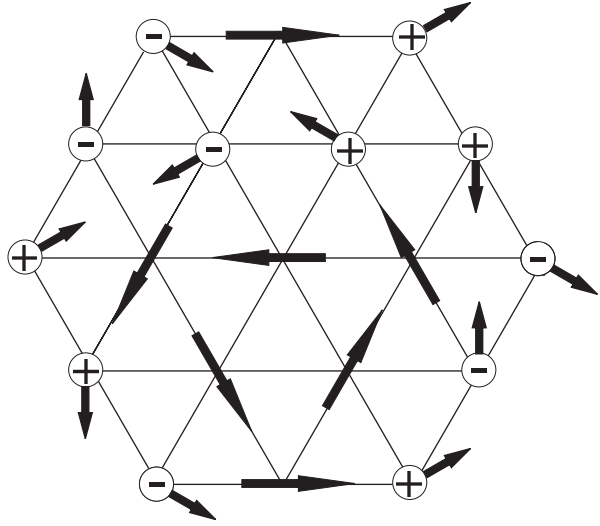}}
\caption{
Expected magnetic ordering in undoped CoO$_2$ plane in the limit of 
a small trigonal field. Shown is the magnetic unit cell which contains 
twelve different lattice sites. Circles with $\pm$ sign indicate the 
out-of-plane component of the magnetic moment canted away from the plane 
by an angle $\theta$ (with $\tan\theta=\pm\sqrt 2$). Associated Bragg spots 
in a momentum space are located at $K/2=(2\pi/3,0)$ and equivalent 
points. (No Bragg intensity at $K$ points). Note that correlations 
in a majority of the bonds are more "ferromagnetic" rather than
"antiferromagnetic".       
}
\label{fig5}
\end{figure}

{\it Small trigonal field}, $\Delta\ll \lambda$.--- The pseudospin 
Hamiltonian has the most symmetric form when the quantization 
axes are along the Co-O bonds. For the Co-Co pairs 1-2, 2-3 and 3-1  
[Fig.\ref{fig4}(b)], respectively, the interactions read as follows:  
\begin{eqnarray}
\label{Hcubic} 
H(1-2)&=&J_{eff}(-S^x_iS^x_j-S^y_iS^y_j+S^z_iS^z_j), \\ \nonumber
H(2-3)&=&J_{eff}(-S^y_iS^y_j-S^z_iS^z_j+S^x_iS^x_j), \\ \nonumber
H(3-1)&=&J_{eff}(-S^z_iS^z_j-S^x_iS^x_j+S^y_iS^y_j),
\end{eqnarray} 
where $J_{eff}=2(B+C)/9\sim 10-15$~meV. Here, we denoted the $f$-pseudospin 
simply by $\vec S$. Interestingly, the $A$-term (\ref{A}) does not 
contribute to the $f$-level interactions in this limit. The projected 
exchange interaction is anisotropic and also strongly depends on the 
bond direction, reflecting the spin-orbital 
mixed character of the ground state wave function, see Eq.(\ref{eq2}). 
Alternation of the antiferromagnetic component from the bond to bond, 
superimposed on the frustrated nature of a triangular lattice, makes the 
Hamiltonian (\ref{Hcubic}) rather nontrivial and interesting. Surprisingly, 
one can gauge away the minus signs in Eqs.(\ref{Hcubic}) in all the 
bonds {\it simultaneously}. To this end, we introduce {\it four} triangular 
sublattices, each having a doubled lattice parameter 2$a$.   
The first sublattice includes the origin, while the others are shifted 
by vectors $\vec{\delta}_{12}$,  $\vec{\delta}_{23}$ and $\vec{\delta}_{31}$ 
[1,2,3 refer to the sites shown in Fig.\ref{fig4}(b)]. Next, we introduce 
a fictitious spin on each sublattice (except for the first one), 
${\vec S}'$, which are obtained from the original $f$-pseudospins $\vec S$ 
by changing the sign of two appropriate components, depending 
on sublattice index [this is a similar trick, used in the context 
of the orbital Hamiltonian (\ref{ytio3}) for ferromagnetic YTiO$_3$; 
see for details Refs.\cite{Kha02,Kha03}]. After these transformations, 
we arrive at very simple result for the fictitious spins: 
$H=J_{eff}({\vec S}'_i \cdot {\vec S}'_j)$ in {\it all the bonds}. 
Thus, the known results \cite{Miy92,Cap99} for the AF-Heisenberg model 
on triangular lattice can be used. Therefore, we take 
120$^{\circ}-$ordering pattern for the fictitious spins and map it back 
to the original spin space. The resulting magnetic order has a large unit cell
shown in Fig.\ref{fig5}. [Four sublattices have been introduced to map the 
model on a fictitious spin space; yet each sublattice contains three 
different orientations of ${\vec S}'$]. Ferro- and antiferromagnetic 
correlations are mixed up in this structure, and the first ones 
are more pronounced as expected from Eq.(\ref{Hcubic}). 
Magnetic order is highly noncollinear and also noncomplanar, 
and a condensation of the spin vorticity in the ground state is apparent. 
The corresponding Bragg peaks are obtained at positions $K/2=(2\pi/3,0)$, 
that is, half-way from the ferromagnetic $\Gamma-$point to the 
AF-Heisenberg $K-$point. Because of the "hidden" symmetry (which becomes 
explicit and takes the form of a global {\it SU(2)} for the fictitious 
spins ${\vec S}'$), the moments can be rotated at no energy cost. 
[Note that the "rotation rules" for the real moments are not that simple 
as for ${\vec S}'$: they are obtained from a global {\it SU(2)} rotation 
of ${\vec S}'$ via the mapping ${\vec S}\Longleftrightarrow{\vec S}'$]. 
Thus, the excitation spectrum is gapless (at $\Delta=0$), 
and can in fact be obtained by folding of the spinwave dispersions 
of the AF-Heisenberg model. We find nodes at the Bragg points 
$K/2$ (and also at $\Gamma$, $K$, $M$ points, but magnetic 
excitations have a vanishing intensity at this points). 
Spin-wave velocity is $v=(3\sqrt3/2)J_{eff}$. A doped-hole motion in such 
spin background should be highly nontrivial --- an interesting problem for 
a future study. 

{\it Large trigonal field}, $\Delta\gg \lambda$.--- In this 
case, a natural quantization axes are the in-plane ($ab$) and 
out-of-plane ($c$) directions. The low energy magnetic Hamiltonian 
obtained reads as follows:  
\begin{equation}
H(\Delta\gg \lambda)=J_cS^z_iS^z_j+J_{ab}(S^x_iS^x_j+S^y_iS^y_j), 
\end{equation}
with $J_c=[A-2(3R-1)B]/9$ and $J_{ab}=(A-B)/9$. As $A\sim B$ and $R>1$, 
we have a large negative $J_c$ and small $J_{ab}$ (ferromagnetic 
Ising-like case). Thus, the {\it large compression} of CoO$_2$ plane 
stabilizes a uniaxial {\it ferromagnetic state} with moments aligned 
along the $c$ direction, and the excitations have a gap. When the AF 
coupling between the different planes is included, expected magnetic 
structure is then of $A$-type (ferro-CoO$_2$ planes coupled 
antiferromagnetically).

The above results show that the nature of magnetic correlations is highly 
sensitive to the trigonal distortion. This is because the variation of  
the ratio $\lambda/\Delta$ strongly modifies the wave-function of 
the lowest Kramers doublet, which determines the anisotropy of intersite 
exchange interactions. In this context, it is tempting to mention 
recent NMR-reports on the observation of magnetic correlations in 
superconducting cobaltates at wave vectors "in between" the ferromagnetic 
and AF 120--degree modes \cite{Nin05}, while correlations change towards the 
ferromagnetic point in samples with a larger trigonal splitting \cite{Iha05}. 
These observations can naturally be understood within the theory presented 
here, assuming that the Na-doped compounds still "remember" a local 
magnetic interactions that we obtained here for the insulating limit. 

{\it Pseudofermion pairing}.--- Another interesting point is that unusual 
superexchange interactions in CoO$_2$ 
layer may have important consequences for the pairing symmetry 
in doped compounds, as argued recently in Ref.\cite{Kha04b}.  
Because of the non-Heisenberg form of the effective $J-$Hamiltonian, 
usual singlet/triplet classification is not very appropriate. 
To visualize a symmetry of two paired spins --- in the spirit of 
the RVB picture ---  we represent the exchange Hamiltonian (\ref{Hcubic}) 
in terms of $f$-fermionic spinons corresponding to the lowest Kramers doublet. 
Choosing now quantization along $z\parallel c$, we find    
\begin{equation}
\label{pair}
H_{ij}=-J_{eff}\Delta_{ij}^{(\gamma)\dagger}\Delta_{ij}^{(\gamma)}, 
\;\;\; 
\Delta_{ij}^{(\gamma)}=(t_{ij,0}+e^{i\phi^{(\gamma)}}t_{ij,1}+ 
e^{-i\phi^{(\gamma)}}t_{ij,-1})/\sqrt3~.   
\end{equation}
Here, $t_{ij,0}=i(f_{i\bar\uparrow}f_{j\bar\downarrow}+
f_{i\bar\downarrow}f_{j\bar\uparrow})/\sqrt{2}$~, 
$t_{ij,1}=f_{i\bar\uparrow}f_{j\bar\uparrow}$ and 
$t_{ij,-1}=f_{i\bar\downarrow}f_{j\bar\downarrow}$ correspond to 
different projections $M=0,\pm 1$ of the total pseudospin $S_{tot}=1$ 
of the pair. The phase $\phi^{(\gamma)}$ in (\ref{pair}) depends on the 
$\langle ij \rangle$-bond direction:  
$\phi^{(\gamma)}=(0,\phi,-\phi)$ for $(12,23,13)$-bonds 
[see Fig.\ref{fig4}(b)], respectively. 
As is evident from Eq.(\ref{pair}), the pairing field 
$\Delta_{ij}^{(\gamma)}$ is {\it spin symmetric but nondegenerate}, 
because it is made of a particular linear combination of $M=0,\pm 1$ 
components, and the total spin of the pair is in fact quenched 
in the ground state. The absence of degeneracy is due 
to the fact that the pairing $J-$interaction has no rotational 
{\it SU(2)} symmetry. In a sense of spin non-degeneracy, the pair 
is in a singlet state, although its wavefunction is composed 
of $M=0,\pm 1$ states and thus {\it spin symmetric}. 
[For the {\it fictitious} spins introduced above, this would however 
read antisymmetric]. When such unusual pairs do condense, a momentum space 
wave-function must be of {\it odd symmetry} (of $p,f,..$-wave character; 
a precise form of the gap functions is given in Ref.\cite{Kha04b}). 
It is interesting to note, that the magnetic susceptibility 
(and hence the NMR Knight shift) is {\it partially} suppressed in 
{\it all three directions} in this paired state. [We recall that $\vec S$ 
represents a total moment of the lowest Kramers level; but 
no Van Vleck contribution is considered here]. 
The relative weights of different $M$-components, which control 
the Knight-shift anisotropy, depend on trigonal distortion 
via the ratio $\Delta/\lambda$, see for details Ref.\cite{Kha04b}.  

\subsection{Spin/orbital polarons in a doped NaCoO$_2$}
Finally, we discuss an interesting consequence of the orbital 
degeneracy in sodium-rich compounds Na$_{1-x}$CoO$_2$ at 
small $x$. They are strongly correlated metals and also show 
magnetic order of $A$-type, which seems surprising in view of the fact 
that only a small number $x$ of magnetic Co$^{4+}$ ions are present (in fact, 
magnetism disappears for large $x$).    
The parent compound NaCoO$_2$ is usually regarded as a band insulator, 
as Co$^{3+}$ ions have a spinless configuration $t_{2g}^6$. However, 
one should note that this state results from a delicate balance 
between the $e_g$-$t_{2g}$ 
crystal field splitting and a strong intraatomic Hund's interaction, 
and a control parameter $10Dq-2J_H$ may change sign even under 
relatively weak perturbations. This would stabilize 
either $t_{2g}^5e_g$ or $t_{2g}^4e_g^2$  
magnetic configurations,  
bringing thereby Mott physics "hidden" in the ground state configuration. 
Doping of NaCoO$_2$ (by removing Na) is a very efficient way of doing it: 
a doped charge on Co$^{4+}$ sites breaks locally the cubic symmetry and 
hence splits the 
$e_g$ and $t_{2g}$ levels on neighboring Co$^{3+}$ sites. This reduces 
the gap between the lowest level of the split $e_g$ doublet and the upper
level of the $t_{2g}$ triplet, favoring a $t_{2g}^5e_g$ (S=1)  
configuration~\cite{Ber04}. As a result, a doped 
charge Co$^{4+}$ is dressed by a hexagon of spin-and-orbitally  
polarized Co$^{3+}$ ions, forming a local
object which can be termed as a spin/orbital polaron. The idea of  
orbital polarons has been proposed and considered in detail 
in Ref.~\cite{Kil99} in the context of weakly doped LaMnO$_3$. The 
$e_g$ level splitting on sites next to a doped hole has been estimated 
to be as large as $\sim$~0.6~eV (in a perovskite structure with 
180$^{\circ}$-bonds), leading to a large binding energies and explaining 
the insulating nature of weakly doped ferromagnetic manganites.      

In fact, doping induced spin-state transmutations have been observed in
perovskite LaCoO$_3$~\cite{Yam96}, in which a small amount of Sr impurities 
triggers magnetism. Because of their nontrivial magnetic response 
to the doping, we may classify NaCoO$_2$ and LaCoO$_3$ as Mott insulators 
with {\it incipient} magnetism. Indeed, a ground state 
with a filled $t_{2g}$ shell looks formally similar to that 
of band insulator but is qualitatively different from the latter: 
NaCoO$_2$ and LaCoO$_3$ have low-lying magnetic states. In LaCoO$_3$, 
they are just $\sim 10$~meV above the ground state. 
Thus, the spin and charge energy scales are completely different.  
In fact, Co$^{3+}$ ions in LaCoO$_3$ fully retain 
their atomic-like multiplet structure, as it is well documented by 
ESR~\cite{Nog02} and inelastic neutron scattering~\cite{Pod05} measurements. 
Spin states of such a Mott insulator can easily be activated 
by doping, temperature {\it etc}. 

The next important point is that the internal spin structure of 
the spin/orbital polarons in NaCoO$_2$ is very different from that 
in perovskites LaCoO$_3$ and LaMnO$_3$. In the latter cases, polarons 
have a large spin due to internal motion of the bare hole within the polaron 
({\it local} double-exchange process) \cite{Kil99}. We argue now that 
{\it due to} 90$^{\circ}$-{\it geometry} of Co-O-Co bonds in NaCoO$_2$, 
the exchange interactions within a polaron are strongly antiferromagnetic, 
and thus a {\it polaron has a total spin one-half} only. 
({\it i}) Consider first a superexchange between Co$^{3+}$ ions 
that surround a central Co$^{4+}$ and form a hexagon. They are    
coupled to each other via the 90$^{\circ}$-superexchange. An antiferromagnetic 
interaction $\tilde J({\vec S}_i \cdot {\vec S}_j)$ between 
two neighboring Co$^{3+}$ spins (S=1) is mediated 
by virtual hoppings of electron between $t_{2g}$ and $e_g$ 
orbitals, $\tilde t=t_{\sigma}t_{\pi}/\Delta_{pd}$, and we find  
$\tilde J\sim\tilde t^2/E$, where $E$ is a relevant charge excitation energy 
(order of $U_d$ or $\Delta_{pd}+U_p/2$). Note that there will be also 
a weaker contribution from $t_{2g}-t_{2g}$ hoppings 
(antiferromagnetic again) and some ferromagnetic corrections 
from the Hund's interaction ($\propto J_H/U$), which are neglected.    
Considering $\tilde t\sim 0.2$~eV and $E\sim 3$~eV, we estimate 
$\tilde J$ could be as large as 10--15~meV. Below this energy scale, spins S=1 
of Co$^{3+}-$hexagon form a singlet and are thus "hidden", but 
they do contribute to the high-temperature magnetic susceptibility in a form 
$C/(T-\theta)$ with a negative $\theta\sim -\tilde J$.         
({\it ii}) Next, one may think that the double exchange process between 
the $t_{2g}$ S=1/2 of the central Co$^{4+}$ and S=1 of the Co$^{3+}$ would 
stabilize a large-spin polaron, but this is not the case. Surprisingly, 
this interaction is also of antiferromagnetic character, because there 
is no direct $e_g-e_g$ transfer (which gives strong ferromagnetic polarons 
in perovskites with 180$^{\circ}$-bonds like LaCoO$_3$ and LaMnO$_3$). 
Instead, there are two nearly equal double-exchange contributions 
which favor different spin alignment: 
(A) an $e_g$ electron of Co$^{3+}$ goes to (a singly occupied) $t_{2g}$ level 
of Co$^{4+}$. This process is possible when the Co$^{3+}$--Co$^{4+}$ pair 
is antiferromagnetic. Another process is that (B) a $t_{2g}$ electron 
of Co$^{3+}$ goes to an empty $e_g$ level of Co$^{4+}$, which favors 
ferromagnetic bond via the Hund's coupling as for the usual double exchange. 
The hopping amplitude in both cases is just the same, $\tilde t$, but 
the kinetic energy gain is slightly larger for the antiferromagnetic 
configuration, because an extra energy $\sim 10Dq$ (somewhat blocking 
a charge transfer) is involved in the second process.    
Therefore, the total double exchange within the pair Co$^{3+}$--Co$^{4+}$ 
is {\it weakly antiferromagnetic} (a small fraction of $\tilde t$). 
We thus expect that each polaron brings about a free spin one-half in total, 
which contributes to the low-energy spin ordering.   
However, a polaron has an {\it internal} excitations to a multitude of its 
larger-spin states which, in principle, could be observed by neutron 
scattering as a nondispersive and broad magnetic modes (in addition 
to a propagating excitations in a spin one-half polaron sector). 

Our overall picture for Na$_{1-x}$CoO$_2$ at small $x$ is that 
of a heavy spin/orbital polaron liquid. When two polarons overlap, 
their spins one-half are coupled {\it ferromagnetically} via 
the mutual polarization of their spin clouds (Co$^{3+}-$hexagons). Therefore, 
the polaron liquid may develop (itinerant) ferromagnetism within the  
CoO$_2$ planes at low temperatures. 
The polarons may form clusters or charge-ordered patterns  
(near commensurate fillings) as discussed in Ref.\cite{Ber04}. 
However, the polaron picture breaks down at large doping $x$, 
and no magnetism is expected therefore in sodium-poor compounds. 
In other words, a heavy-fermion behavior and magnetism of weakly 
doped NaCoO$_2$ originate from the spin-state transmutation of Co$^{3+}$ 
ions near doped holes, resulting in a narrow polaron bands. 
An apparent paradox of a {\it large negative} $\theta$ seen 
in susceptibility, and a {\it small positive} one inferred 
from spinwave dispersions~\cite{Bay05} is a natural consequence 
of our picture: The former one reflects strong AF-couplings (active at 
large T) within the polaron, while the latter stems from a relatively 
weak ferromagnetic couplings between the spin-one-half polarons at low 
energy limit, contributing to magnons. The in-plane coupling subtracted 
from magnons is found to be small (only of the order of the coupling
between the planes) \cite{Bay05,Hel05}. This is because it stems from 
a residual effective interactions between {\it a dilute} system of 
spin-one-half polarons. Our explanation implies that a quasi 
two-dimensional layered structure of cobaltates should show up 
in magnetic excitations of the {\it magnetically dense} compounds 
close to the undoped parent system CoO$_2$. In that limit, we believe 
that $|J_{ab}/J_c|\gg 1$, just like in a closely related, 
but {\it regular} magnet NaNiO$_2$ with the same crystal 
and ($A$-type) magnetic structure, where $|J_{ab}/J_c|\sim 15$ \cite{Lew05}. 
An important remark concerning the magnon gaps: since all the orbital 
levels are well split within the polaron (lifting the degeneracy by
orbital-charge coupling\cite{Kil99}), the magnetic moment of the polaron 
is mostly of the spin origin. Thus, a low-energy magnetic excitations 
of a polaron liquid should be of the Heisenberg-type, as in fact observed 
in sodium-rich samples\cite{Bay05,Hel05}. 

In our view, marked changes in the properties of cobaltates around 
Na$_{0.7}$CoO$_2$ are associated with a collapse of the polaron picture. 
In the magnetic sector, this implies a breakdown of dynamical separation 
on the strong-AF (weak-F) magnetic bonds within (in-between) the polarons, and 
a uniform distribution of the exchange interactions sets in. 

Summarizing results for the magnetic interactions in the two limiting 
cases considered in this section: --- a "pure" CoO$_2$ 
and weakly doped NaCoO$_2$ --- we conclude, that the nature of 
magnetic correlations in cobaltates is very rich and strongly sensitive 
to the composition. Description in terms of a simple AF-Heisenberg models 
with a uniform interactions in all the bonds is not sufficient. 
Depending on the strength of the trigonal distortion and doping level, 
we find the spin structures ranging from a highly nontrivial one 
as shown in Fig.\ref{fig5}, to a conventional $A$-type structures. 
The $A$-type correlations, which would be favored in a sodium-poor sample by 
large trigonal field, have an uniaxial anisotropy. On the other hand, the 
$A$-type state in a sodium-rich compounds is isotropic. Besides a conventional
magnons, this state may reveal an interesting high-energy response 
stemming from its "internal" polaron structure. 

\section{Summary}
We considered several mechanisms lifting the orbital degeneracy, which 
are based on: ({\it i}) electron-lattice coupling, ({\it ii}) the spin-orbital 
superexchange, and ({\it iii}) a relativistic spin-orbit coupling. 
We discussed mostly limiting cases to understand specific features 
of each mechanism.  Reflecting the different nature of underlying forces, 
these mechanisms usually compete, and this may lead to nontrivial 
phase diagrams as discussed for example for YVO$_3$. 
This underlines the important role of the orbital degrees of freedom 
as a sensitive control parameter of the electronic phases 
in transition metal oxides. 

We demonstrated the power of the "orbital-angle" ansatz in manganites, where 
the validity of classical orbital picture is indeed obvious because of large
lattice distortions. In titanites and vanadates with much less distorted
structures, we find that this simple ansatz is not sufficient and the orbitals
have far more freedom in their "angle" dynamics. Here, the lattice distortions 
serve as an important control parameter and are thus essential, 
but the quantum exchange process between the orbitals becomes a "center of
gravity". 

Comparing the $e_g$ and $t_{2g}$ exchange models, we found that the 
former case is more classical: the $e_g$ orbitals are less frustrated 
and the order-from-disorder mechanism is very effective in lifting the 
frustration by opening the orbital gap. In fact, the low-energy fixed 
point, formed below the orbital gap in $e_g$ spin-orbital models, can 
qualitatively be well described in terms of classical "orbital-angle" picture. 
A strong JT nature of the $e_g$ quadrupole makes in reality this 
description just an ideal for all the relevant energy/temperature scales. 

A classical "orbital-angle" approach to the $t_{2g}$ spin-orbital 
model fails in a fatal way. Here, the orbital frustration can only be 
resolved by strong quantum disorder effects, and dynamical coupling 
between the spins and orbitals is at the heart of the problem. 
This novel aspect of the "orbital physics", emerged from recent 
experimental and theoretical studies of pseudocubic titanites, 
provide a key in understanding unusual properties of these materials, 
and make the field as interesting as the physics of quantum spin systems. 
We believe that the ideas developed in this paper should be relevant 
also in other $t_{2g}$ orbital systems, where the quantum orbital 
physics has a chance to survive against orbital-lattice coupling. 

Unusual magnetic orderings in layered cobaltates, stabilized by 
a relativistic spin-orbit coupling are predicted in this paper, 
illustrating the importance of this coupling in compounds based 
on late-3$d$ and 4$d$ ions. Finally, we considered how the orbital 
degeneracy can be lifted locally around doped holes, resulting 
in a formation of spin/orbital polarons in weakly doped 
NaCoO$_2$. The idea of a dilute polaron liquid provides here a coherent 
understanding of otherwise puzzling magnetic properties. 

In a broader context, a particular behavour of orbitals in the  
reference insulating compounds should have an important consequences for 
the orbital-related features of metal-insulator transitions under 
pressure/temperature/composition, {\it e.g.} whether they are 
"orbital-selective" \cite{Kog05} or not. In case of LaTiO$_3$, where 
the three-band physics is well present already in the insulating state, 
it seems natural not to expect the orbital selection. On the other hand, 
the orbital selection by superexchange and/or lattice interactions 
is a common case in other insulators ({\it e.g.}, $xy$ orbital selection 
in vanadates); here, we believe that the "dimensionality reduction" 
phenomenon --- which is so apparent in the "insulating" models we 
discussed --- should show up already in a metallic side, {\it partially} 
lifting the orbital degeneracy and hence supporting the
orbital-selective transition picture. 

\section*{Acknowledgements}
I would like to thank B.~Keimer and his group members for many stimulating
discussions, K.~Held and O.K.~Andersen for the critical reading of 
the manuscript and useful comments. 
This paper benefited a lot from our previous work on orbital physics, 
and I would like to thank all my collaborators on this topic. 



\begin{thebibliography}{99}
\bibitem{Ima98} M.~Imada, A.~Fujimori, and Y.~Tokura,
                Rev.~Mod.~Phys. {\bf 70}, 1039 (1998).

\bibitem{Tok00} Y.~Tokura and N.~Nagaosa, 
                Science {\bf 288}, 462 (2000).

\bibitem{Mae04} S.~Maekawa, T.~Tohyama, S.E.~Barnes, S.~Ishihara,
		W.~Koshibae, and G.~Khaliullin,  
                {\it Physics of Transition Metal Oxides},  
		Springer Series in Solid State Sciences, vol. 144    
                (Springer-Verlag, Berlin, 2004).

\bibitem{Goo63} J.B.~Goodenough, {\it Magnetism and Chemical Bond} 
                (Interscience Publ., New York-London, 1963).

\bibitem{Kan59} J.~Kanamori, J.~Phys.~Chem.~Solids {\bf 10}, 87 (1959); 
                J.~Appl.~Phys. {\bf 31}, S14 (1960).

\bibitem{Kug82} K.I.~Kugel and D.I.~Khomskii, 
                Sov.~Phys.~Usp. {\bf 25}, 231 (1982). 

\bibitem{Kha00} G.~Khaliullin and S.~Maekawa, 
                Phys.~Rev.~Lett. {\bf 85}, 3950 (2000).

\bibitem{Kha01a} G.~Khaliullin, Phys.~Rev.~B {\bf 64}, 212405 (2001).

\bibitem{Goo04} For a recent review of the physical properties of perovskites, 
                see J.B.~Goodenough, Rep.~Prog.~Phys. {\bf 67}, 1915 (2004).

\bibitem{note1} $G$-type AF structure: spins are staggered in all three 
                directions. $C$-type ($A$-type) structure consists 
                of ferromagnetic chains (planes) with AF order between them.  

\bibitem{Gri61} J.S.~Griffith, {\it The Theory of Transition-Metal Ions} 
                (Cambridge University Press, Cambridge 1961).

\bibitem{Kov04} N.N.~Kovaleva, A.V.~Boris, C.~Bernhard, A.~Kulakov, 
                A.~Pimenov, A.M.~Balbashov, G.~Khaliullin, and B.~Keimer,
                Phys.~Rev.~Lett. {\bf 93}, 147204 (2004).

\bibitem{Mil96} A.J.~Millis, Phys.~Rev.~B {\bf 53}, 8434 (1996).

\bibitem{Fei99} L.F.~Feiner and A.M.~Ole\'s, 
                Phys.~Rev.~B {\bf 59}, 3295 (1999).
 
\bibitem{Kha04a} G.~Khaliullin, P.~Horsch, and A.M.~Ole\'s, 
                 Phys.~Rev.~B {\bf 70}, 195103 (2004).

\bibitem{Hir96} K.~Hirota, N.~Kaneko, A.~Nishizawa, and Y.~Endoh, 
                J.~Phys.~Soc.~Jpn. {\bf 65}, 3736 (1996). 
                $J_c=1.21\pm 0.05$~meV, $J_{ab}=-1.67\pm 0.02$~meV. 

\bibitem{Loa01} I.~Loa, P.~Adler, A.~Grzechnik, K.~Syassen, U.~Schwarz, 
                M.~Hanfland, G.Kh.~Rozenberg, P.~Gorodetsky, and
                M.P.~Pasternak, Phys.~Rev.~Lett. {\bf 87}, 125501 (2001).   

\bibitem{Kat97} T.~Katsufuji, Y.~Taguchi, and Y.~Tokura, 
                Phys.~Rev.~B {\bf 56}, 10145 (1997).

\bibitem{Kei00} B.~Keimer, D.~Casa, A.~Ivanov, J.W.~Lynn, M.v.~Zimmermann, 
                J.P.~Hill, D.~Gibbs, Y.~Taguchi, and Y.~Tokura,  
                Phys.~Rev.~Lett. {\bf 85}, 3946 (2000).

\bibitem{Ulr02} C.~Ulrich, G.~Khaliullin, S.~Okamoto, M.~Reehuis, A.~Ivanov,
                H.~He, Y.~Taguchi, Y.~Tokura, and B.~Keimer, 
                Phys.~Rev.~Lett. {\bf 89}, 167202 (2002).

\bibitem{Cwi03} M.~Cwik, T.~Lorenz, J.~Baier, R.~M\"uller, G.~Andr\'e, 
                F.~Bour\'ee, F.~Lichtenberg, A.~Freimuth, R.~Schmitz, 
                E.~M\"uller--Hartmann, and M.~Braden,
                Phys.~Rev.~B {\bf 68}, 060401 (2003).

\bibitem{Kha02} G.~Khaliullin and S.~Okamoto,
                Phys.~Rev.~Lett. {\bf 89}, 167201 (2002).

\bibitem{Kha03} G.~Khaliullin and S.~Okamoto, 
                Phys.~Rev.~B {\bf 68}, 205109 (2003).

\bibitem{Moc03} M.~Mochizuki and M.~Imada, 
                Phys.~Rev.~Lett. {\bf 91}, 167203 (2003).

\bibitem{Pav04} E.~Pavarini, S.~Biermann, A.~Poteryaev, A.I.~Lichtenstein, 
                A.~Georges, and O.K.~Andersen, 
                Phys.~Rev.~Lett. {\bf 92}, 176403 (2004). 

\bibitem{Miz96} T.~Mizokawa and A.~Fujimori,
                Phys.~Rev.~B {\bf 54}, 5368 (1996).

\bibitem{Saw97} H.~Sawada, N.~Hamada, and K.~Terakura,
                Physica B {\bf 237-238}, 46 (1997).

\bibitem{Aki01} J.~Akimitsu, H.~Ichikawa, N.~Eguchi, T.~Miyano, M.~Nishi, 
                and K.~Kakurai, J.~Phys.~Soc.~Jpn. {\bf 70}, 3475 (2001).

\bibitem{Kiy03} T.~Kiyama and M.~Itoh, 
                Phys.~Rev.~Lett. {\bf 91}, 167202 (2003).

\bibitem{Kiy05} T.~Kiyama, H.~Saitoh, M.~Itoh, K.~Kodama, 
                H.~Ichikawa, and J.~Akimitsu,  
                J.~Phys.~Soc.~Jpn. {\bf 74}, 1123 (2005).

\bibitem{Kub05} In fact, the resonant x-ray scattering data shows that 
                the orbital polarization in LaTiO$_3$ is much weaker 
                than in YTiO$_3$: 
                M.~Kubota, H.~Nakao, Y.~Murakami, Y.~Taguchi, M.~Iwama, and 
                Y.~Tokura,  Phys.~Rev.~B {\bf 70}, 245125 (2005).

\bibitem{Pav05} E.~Pavarini, A.~Yamasaki, J.~Nuss, and O.K.~Andersen, 
                New J.~Phys. {\bf 7}, 188 (2005). 

\bibitem{Sol04} I.V.~Solovyev, Phys.~Rev.~B {\bf 69}, 134403 (2004).

\bibitem{Kel04} G.~Keller, K.~Held, V.~Eyert, D.~Vollhardt, and V.I.~Anisimov, 
                Phys.~Rev.~B {\bf 70}, 205116 (2004).

\bibitem{Nel67} E.D.~Nelson, J.Y.~Wong, and A.L.~Schawlow, 
                Phys.~Rev. {\bf 156}, 298 (1967).

\bibitem{Ulr05}  C.~Ulrich, A.~G\"ossling, M.~Gr\"uninger, M.~Guennou,
                 H.~Roth, M.~Cwik, T.~Lorenz, G.~Khaliullin, and B.~Keimer, 
                 cond-mat/0503106. 

\bibitem{Shi95} Z.-P.~Shi and R.R.P.~Singh,
                Phys.~Rev.~B {\bf 52}, 9620 (1995). 

\bibitem{Miy00} S.~Miyasaka, T.~Okuda, and Y.~Tokura, 
                Phys.~Rev.~Lett. {\bf 85}, 5388 (2000). 

\bibitem{Kha01b} G.~Khaliullin, P.~Horsch, and A.M.~Ole\'s,
                Phys.~Rev.~Lett. {\bf 86}, 3879 (2001). 

\bibitem{Ulr03} C.~Ulrich, G.~Khaliullin, J.~Sirker, M.~Reehuis, M.~Ohl,
                S.~Miyasaka, Y.~Tokura, and B.~Keimer, 
                Phys.~Rev.~Lett. {\bf 91}, 257202 (2003).

\bibitem{Sir03} J.~Sirker and G.~Khaliullin, 
                Phys.~Rev.~B {\bf 67}, 100408(R) (2003).

\bibitem{Yan04} J.-Q.~Yan, J.-S.~Zhou, and J.B.~Goodenough, 
                Phys.~Rev.~Lett. {\bf 93}, 235901 (2004). 

\bibitem{Kha06} G.~Khaliullin, (unpublished). The predicted spin couplings 
                for the ferrates are: NN $J_1\simeq -2.2$~meV, second NN 
                $J_2\simeq 0.3$~meV, and fourth NN $J_4\simeq 0.5$~meV.  

\bibitem{Leb04} A.~Lebon, P.~Adler, C.~Bernhard, A.~Boris, A.~Pimenov, 
                A.~Maljuk, C.T.~Lin, C.~Ulrich, and B.~Keimer, 
                Phys.~Rev.~Lett. {\bf 92}, 037202 (2004).

\bibitem{Bri99} J.~van~den~Brink, P.~Horsch, F.~Mack, and A.M.~Ole\'s, 
                Phys.~Rev.~B {\bf 59}, 6795 (1999). 

\bibitem{Tsv95} For a discussion of the order-from-disorder phenomena 
                in frustrated spin systems, see A.~M.~Tsvelik,
                {\it Quantum Field Theory in Condensed Matter Physics}
                (Cambridge University Press, Cambridge, 1995), 
                Chap.~17, and references therein. 

\bibitem{Kub02} K.~Kubo, J.~Phys.~Soc.~Jpn. {\bf 71}, 1308 (2002).

\bibitem{Kug75} K.I.~Kugel and D.I.~Khomskii, 
                Sov.~Phys.~Solid~State {\bf 17}, 285 (1975).

\bibitem{Oki95} Y.~Okimoto, T.~Katsufuji, Y.~Okada, T.~Arima, and Y.~Tokura,
                Phys.~Rev.~B {\bf 51}, 9581 (1995). 

\bibitem{Ruc05} R.~R\"uckamp, E.~Benckiser, M.W.~Haverkort, H.~Roth, 
                T.~Lorenz, A.~Freimuth, L.~Jongen, A.~M\"oller, G.~Meyer, 
                P.~Reutler, B.~B\"uchner, A.~Revcolevschi, S.-W. Cheong, 
                C.~Sekar, G.~Krabbes, and M.~Gr\"uninger, 
                New J.~Phys. {\bf 7}, 144 (2005).  

\bibitem{Ish04} S.~Ishihara, Phys.~Rev.~B {\bf 69}, 075118 (2004). 

\bibitem{Fei97} L.F.~Feiner, A.M.~Ole\'s,  and J.~Zaanen,
                Phys.~Rev.~Lett. {\bf 78}, 2799 (1997).

\bibitem{Kha97} G.~Khaliullin and V.~Oudovenko, 
                Phys.~Rev.~B {\bf 56}, R14243 (1997).

\bibitem{Tsu03} H.~Tsunetsugu and Y.~Motome, 
                Phys.~Rev.~B {\bf 68}, 060405(R) (2003).

\bibitem{Har03} A.B.~Harris, T.~Yildirim, A.~Aharony, 
                O.~Entin-Wohlman, and I.Ya.~Korenblit, 
                Phys.~Rev.~Lett. {\bf 91}, 087206 (2003).  

\bibitem{Li98} Y.Q.~Li, M.~Ma, D.N.~Shi, and F.C.~Zhang, 
               Phys.~Rev.~Lett. {\bf 81}, 3527 (1998).

\bibitem{Fri99} B.~Frischmuth, F.~Mila, and M.~Troyer, 
                Phys.~Rev.~Lett. {\bf 82}, 835 (1999). 

\bibitem{Sch96} H.J.~Schulz, Phys.~Rev.~Lett. {\bf 77}, 2790 (1996).

\bibitem{Moc01} M.~Mochizuki and M.~Imada, 
                J.~Phys.~Soc.~Jpn. {\bf 70}, 1777 (2001).

\bibitem{Miy02} S.~Miyasaka, Y.~Okimoto, and Y.~Tokura,
                J.~Phys.~Soc.~Jpn. {\bf 71}, 2086 (2002).

\bibitem{Mot03} Y.~Motome, H.~Seo, Z.~Fang, and N.~Nagaosa, 
                Phys.~Rev.~Lett. {\bf 90}, 146602 (2003). 

\bibitem{Miz99} T.~Mizokawa, D.I.~Khomskii, and G.A.~Sawatzky, 
                Phys.~Rev.~B {\bf 60}, 7309 (1999).

\bibitem{She02} Shun-Qing Shen, X.C.~Xie, and F.C.~Zhang, 
                Phys.~Rev.~Lett. {\bf 88}, 027201 (2002). 

\bibitem{Hor03} P.~Horsch, A.M.~Ole\'s, and G.~Khaliullin, 
                Phys.~Rev.~Lett. {\bf 91}, 257203 (2003). 

\bibitem{Miy04} S.~Miyashita, A.~Kawaguchi, N.~Kawakami, and G.~Khaliullin, 
                Phys.~Rev.~B {\bf 69}, 104425 (2004).

\bibitem{Fuj05} J.~Fujioka, S.~Miyasaka, and Y.~Tokura, 
                Phys.~Rev.~B {\bf 72}, 024460 (2005). 
 
\bibitem{Ish05} S.~Ishihara, Phys.~Rev.~Lett. {\bf 94}, 156408 (2005).  

\bibitem{Oga90} M.~Ogata and H.~Shiba, Phys.~Rev.~B {\bf 41}, 2326 (1990).

\bibitem{Kan56} J.~Kanamori, Prog.~Theor.~Phys. {\bf 17}, 177; 197 (1956).  

\bibitem{Hol71} T.M.~Holden, W.J.L.~Buyers, E.C.~Svensson, R.A.~Cowley, 
                M.T.~Hutchings, D.~Hukin, and R.W.H.~Stevenson, 
                J.~Phys.~C: Solid State Phys. {\bf 4}, 2127 (1971).

\bibitem{Foo03} M.L.~Foo, Y.~Wang, S.~Watauchi, H.W.~Zandbergen, T.~He, 
                R.J.~Cava, and N.P.~Ong, 
                Phys.~Rev.~Lett. {\bf 92}, 247001 (2004).   

\bibitem{Lee05} K.-W.~Lee and W.E.~Pickett, 
                Phys.~Rev.~B {\bf 72}, 115110 (2005).

\bibitem{Kos03} W.~Koshibae and S.~Maekawa, 
                Phys.~Rev.~Lett. {\bf 91}, 257003 (2003). 

\bibitem{Fig00} B.N.~Figgis and M.A.~Hitchman, 
                {\it Ligand Field Theory and Its Applications} 
                (New York: Wiley-VCH, 2000).

\bibitem{Kha04b} G.~Khaliullin, W.~Koshibae, and S.~Maekawa,  
                 Phys.~Rev.~Lett. {\bf 93}, 176401 (2004). 

\bibitem{Zaa85} J.~Zaanen, G.A.~Sawatzky, and J.W.~Allen, 
                Phys.~Rev.~Lett. {\bf 55}, 418 (1985).

\bibitem{Miy92} S.J.~Miyake, J.~Phys.~Soc.~Jpn. {\bf 61}, 983 (1992).  

\bibitem{Cap99} L.~Capriotti, A.E.~Trumper, and S.~Sorella, 
                Phys.~Rev.~Lett. {\bf 82}, 3899 (1999).

\bibitem{Nin05} F.L.~Ning and T.~Imai, 
                Phys.~Rev.~Lett. {\bf 94}, 227004 (2005).  

\bibitem{Iha05} Y.~Ihara, K.~Ishida, C.~Michioka, M.~Kato, K.~Yoshimura, 
                K.~Takada, T.~Sasaki, H.~Sakurai, and E.~Takayama-Muromachi, 
                J.~Phys.~Soc.~Jpn. {\bf 74}, 867, (2005). 

\bibitem{Ber04} C.~Bernhard, A.V.~Boris, N.N.~Kovaleva, G.~Khaliullin, 
                A.V.~Pimenov, L.~Yu, D.P.~Chen, C.T.~Lin, and B.~Keimer, 
                Phys.~Rev.~Lett. {\bf 93}, 167003 (2004).

\bibitem{Kil99} R.~Kilian and G.~Khaliullin, 
                Phys.~Rev.~B {\bf 60}, 13458, (1999). 

\bibitem{Yam96} S.~Yamaguchi, Y.~Okimoto, H.~Taniguchi, and Y.~Tokura,  
                Phys.~Rev.~B {\bf 53}, R2926, (1996).

\bibitem{Nog02} S.~Noguchi, S.~Kawamata, K.~Okuda, H.~Nojiri, and M.~Motokawa, 
                Phys.~Rev.~B {\bf 66}, 094404 (2002). 

\bibitem{Pod05} A.~Podlesnyak, S.~Streule, J.~Mesot, M.~Medarde,
                E.~Pomjakushina, K.~Conder, M.W.~Haverkort, and D.I.~Khomskii, 
                cond-mat/0505344.  

\bibitem{Bay05} S.P.~Bayrakci, I.~Mirebeau, P.~Bourges, Y.~Sidis, M.~Enderle,
                J.~Mesot, D.P.~Chen, C.T.~Lin, and B.~Keimer, 
                Phys.~Rev.~Lett. {\bf 94}, 157205 (2005).  

\bibitem{Hel05} L.M.~Helme, A.T.~Boothroyd, R.~Coldea, D.~Prabhakaran, 
                D.A.~Tennant, A.~Hiess, and J.~Kulda,  
                Phys.~Rev.~Lett. {\bf 94}, 157206 (2005).  

\bibitem{Lew05} M.J.~Lewis, B.D.~Gaulin, L.~Filion, C.~Kallin, A.J.~Berlinsky, 
                H.A.~Dabkowska, Y.~Qiu, and J.R.D.~Copley, 
                Phys.~Rev.~B {\bf 72}, 014408 (2005). 

\bibitem{Kog05} A.~Koga, N.~Kawakami, T.M.~Rice, and M.~Sigrist, 
                Phys.~Rev.~B {\bf 72} 045128 (2005). 

\end{thebibliography}
\end{document}